\documentclass[preprint]{elsarticle}

\makeatletter
\def\ps@pprintTitle{%
  \let\@oddhead\@empty
  \let\@evenhead\@empty
  \let\@oddfoot\@empty
  \let\@evenfoot\@oddfoot
}
\makeatother

\usepackage[utf8]{inputenc}  
\usepackage[T1]{fontenc}     
\usepackage{lmodern}         
\usepackage{microtype}       
\usepackage[scaled]{helvet}  
\usepackage[english]{babel}  
\usepackage{graphicx}        
\usepackage[export]{adjustbox}
\usepackage{subfig}
\usepackage[normalem]{ulem}
\usepackage{{booktabs}}
\usepackage[usenames,dvipsnames]{xcolor}
\usepackage{algorithm}
\usepackage{algorithmic}
\usepackage{eqparbox}

\usepackage{tikz} \usetikzlibrary{calc, arrows.meta, intersections, patterns, positioning, shapes.misc, fadings, through,decorations.pathreplacing}

\usepackage[bookmarks=false]{hyperref} 
\usepackage{comment}         
\biboptions{sort&compress}   

\journal{}

\usepackage{amsmath,amsfonts,amssymb}

\usepackage{xcolor}
\usepackage{todonotes}
\usepackage[toc,page]{appendix}
\usepackage{bm}

\graphicspath{{fig/}}
\usepackage{geometry}
\begin{document}

\begin{frontmatter}

\title{Quadratic unconstrained binary optimization and constraint programming approaches for lattice-based cyclic peptide docking}

\author[add1]{J.~Kyle Brubaker}
\address[add1]{Amazon Advanced Solutions Lab, Seattle, Washington 98170, USA}
\ead{johbruba@amazon.com}

\author[add1]{Kyle E.~C.~Booth}

\author[add3]{Akihiko Arakawa}
\address[add3]{Research Division, Chugai Pharmaceutical Co. Ltd., 216, Totsuka-cho, Totsuka-ku, Yokohama, Kanagawa 244-8602, Japan}

\author[add1]{Fabian Furrer}

\author[add2]{Jayeeta Ghosh}
\address[add2]{AWS Professional Services, Seattle, Washington 98170, USA}

\author[add3]{Tsutomu Sato}

\author[add1]{Helmut G.~Katzgraber}

\begin{abstract}

The peptide-protein docking problem is an important problem in structural biology that facilitates rational and efficient drug design. 
In this work, we explore modeling and solving this problem with the quantum-amenable quadratic unconstrained binary optimization (QUBO) formalism. 
Our work extends recent efforts by incorporating the objectives and constraints associated with peptide cyclization and peptide-protein docking in the two-particle model on a tetrahedral lattice. 
We propose a ``resource efficient'' QUBO encoding for this problem, and baseline its performance with a novel constraint programming (CP) approach. 
We implement an end-to-end framework that enables the evaluation of our methods on instances from the Protein Data Bank (PDB). 
Our results show that the QUBO approach, using a classical simulated annealing solver, is able to find feasible conformations for problems with up to 6 peptide residues and 34 target protein residues, but has trouble scaling beyond this problem size. 
In contrast, the CP approach can solve problems with up to 13 peptide residues and 34 target protein residues. 
We conclude that while QUBO can be used to successfully tackle this problem, its scaling limitations and the strong performance of the CP method suggest that it may not be the best choice.

\end{abstract}
\end{frontmatter}

\section{Introduction}\label{sec:introduction}

Peptide drugs are composed of short sequences of amino acids and are an important therapeutic modality for clinical treatments.
These drugs have the potential to acquire higher binding affinity and selectivity, and to exhibit wider therapeutic windows than small molecule drugs, although there are drawbacks related to pharmacokinetic properties such as membrane permeability and metabolic stability \cite{Wang_2022, Zhang_2022}. 
Recently, it was reported that some pharmacokinetic issues could be resolved by the cyclization of peptides and incorporation of nonproteinogenic amino acids \cite{Ohta_2023, Nomura_2022, Corbett_2021}. 
Therefore, cyclic peptides are increasingly expected to be a promising therapeutic modality for targeting both extracellular molecules, and intracellular biomolecules by oral administration \cite{Tanada_2023, Tucker_2021}. 

Structure-based drug design (SBDD), where a drug is designed based on three-dimensional structures of drug candidates and their target, is now widely applied to small-molecule drug discovery research.
Thanks to advances in structural biology and artificial intelligence (AI), three-dimensional models of various target proteins are readily available for SBDD \cite{Jumper_2021, Lin_2022, Baek_2021, Wu_2022, Ahdritz_2024}. 

In contrast to small molecules, cyclic peptides contain more rotatable bonds and thus have more possible conformations. 
Furthermore, their conformations depend on the surrounding environment (e.g., water or lipid membrane) and the nature of their target \cite{Corbett_2021}. 
Consequently, building practical three-dimensional models of cyclic peptide-target complexes is a challenging task, though there are several docking tools for cyclic peptides tackling this challenge \cite{Geng_2016, Charitou_2022, Alogheli_2017, Zhang_2019, Kosugi_2023, Zhang_2024}.

A promising research direction approximates the conformation prediction problem as a combinatorial search over a fixed lattice \cite{Dill_1985, Lau_1989,Berger_1998}. 
Early versions of this approach model the peptide as a series of particles (points) in space, where each peptide residue is represented by a single particle, and the particles are placed on the vertices of a two-dimensional lattice grid. 
Any pair of particles that are within a fixed distance of each other contribute interaction potential according to whether both members of the pair are considered hydrophobic or hydrophilic (e.g., the HP model) \cite{Camacho_1993}.
Some straightforward extensions of this formulation include: i) a two-particle representation for the peptide residues, with one particle representing the residue's main-chain backbone and the other representing the same residue's side chain, ii) the use of Miyazawa-Jernigan (MJ) potentials as the interaction potentials between residues, and iii) the use of three-dimensional lattices \cite{Abkevich_1994, Pande_2000}.
The task is then to identify the placement of these peptide particles such that the MJ potential is optimized (minimized) while all known constraints, e.g., that no two particles occupy the same lattice vertex, are satisfied.
This placement task is a combinatorial search problem that has been shown to be NP-complete~\cite{Berger_1998,Crescenzi_1998}, imposing a severe limitation on the size of problems that can be solved with a classical computer. 
A potential avenue to overcome these scaling issues in the near term is to use noisy intermediate scale quantum (NISQ) technology such as quantum annealing~\cite{Babbush_2014,Perdomo_2012,Outeiral_2021,Irback_2022}, quantum variational algorithms~\cite{Robert_2021,Boulebnane_2022}, or Gaussian Boson sampling~\cite{Banchi_2020}. 
Common to most of these approaches is the formulation of the problem as a quadratic unconstrained binary optimization (QUBO) problem.
We provide a concise summary of the relevant literature in Appendix \ref{appendix:related_work}.

In this work, we extend the lattice-based conformation search from the literature to incorporate the objectives and constraints associated with peptide cyclization and peptide docking with a target protein. 
We explore a number of QUBO modeling formulations, and based on preliminary model scaling analysis (described briefly in Appendix \ref{appendix:scaling}), elect to extend the ``resource efficient'' turn encoding approach in~\cite{Robert_2021}. 
To compare the QUBO with a classical optimization approach, we also derive a constraint programming (CP) model for the problem. 
We implement an end-to-end framework that allows us to automatically process a problem instance from the Protein Data Bank (PDB) and produce conformation solutions with our QUBO-based and CP approaches \cite{PDB}. 
We conduct an empirical assessment on a set of peptide-protein docked instances from the PDB and conclude that while the QUBO-based approaches can be used to model and solve the problem, they may not be the best-suited formalism. 
In contrast, the CP model is able to find the optimal solutions for all considered problem instances.

\section{Methods} \label{sec:problem}

\subsection{Problem Definition}
\begin{figure}[ht!]
  \centering
  \includegraphics[width=0.4\textwidth]{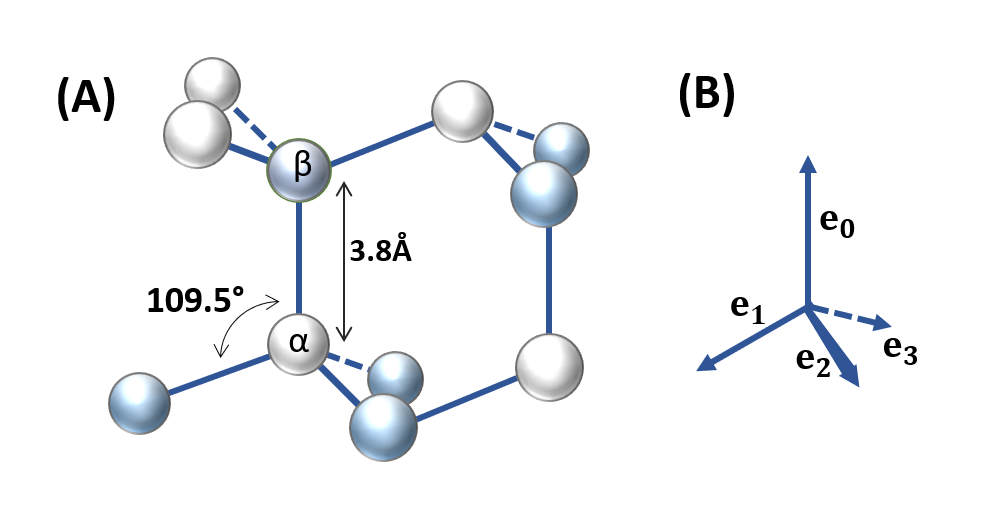}
  \caption{(A) Tetrahedral lattice structure $\mathcal{T}$ with
  vertices on sublattice $\mathcal{T}_\alpha$ indicated with white circles and $\mathcal{T}_\beta$ in blue circles. (B) Turn directions of the tetrahedral lattice.}
  \label{fig:3dtetrahedral}
\end{figure}

We employ a two-particle coarse grain (CG) residue representation, describing main and side-chain particles for all residues except for a specified set of small residues (e.g., Glycine), for which we use a one-particle representation.
We place these peptide particles on a three-dimensional tetrahedral lattice, for instance as depicted in Figure \ref{fig:3dtetrahedral}, defined by the set of its vertices $\mathcal{T}$. 

For a given problem instance, we consider a cyclic peptide consisting of an amino acid sequence $A = (A_1,...,A_N)$, where $A_i$ specifies the particular amino acid type for sequence location $i$, $A_i$ is adjacent to $A_{i+1}$, and $N$ represents the number of amino acids.
We use $A_i$ to refer to the main chain particle and use a superscript $s$ to denote the corresponding side-chain particle $A^s_i$ that is assumed to be adjacent to $A_i$.
Cyclization bonds of the peptide are expressed by pairs of residue numbers $C = \{(i_1,j_1), ... ,(i_K,j_K)\}$ for $K$ specified pairs that are bonded, where typically $K=1$.

In this work, we consider a smaller, more focused area of the external protein, namely the ``active site.'' More details on how the active site is constructed for our experiments are provided in Appendix \ref{appendix:experiments}. 
This external protein active site is represented as $P=\{B_1,...,B_M\}$, where $B_i$ specifies the particular residue (i.e.,  amino acid type) at sequence location $i$ and $M$ is the number of residues.

We model the interaction energy between amino acid residues by Miyazawa-Jernigan (MJ) potentials~\cite{Miyazawa_1985, Miyazawa_1996}, which specify a fixed interaction potential between unique pairs of amino acid types.
For simplicity of notation, we denote $\epsilon_{ij}$ to be the MJ interaction between residues $A_i$ and $A_j$.
For the peptide, we restrict the interaction to non-bonded particles (i.e.,  main- and side-chain particles) that are first nearest neighbours (1-NN) on the lattice (i.e.,  a distance of one lattice edge); pairs of residues that are not 1-NN in the solution do not contribute any interaction energy.

Given the coordinates of the protein residues $P$, the lattice vertices $\mathcal{T_B} = \{t^B_{k}\}_k \subset \mathcal{T}$ that are ``blocked'' by the protein due to steric hindrance can be calculated using a specified blocking radius (e.g., $3.8$\AA). 
Similarly, using a specified interaction radius (e.g., $6.5$\AA), we can compute the lattice vertices $\mathcal{T_I} = \{t^I_{l}\}_{l} \subset \mathcal{T}$ on which the protein residues effect an interaction potential for the peptide. 
Here, we assume that all blocking points $\mathcal{T_B}$ are removed from $\mathcal{T_I}$, as the steric hindrance prohibits peptide residues from being placed on those vertices.
We denote the total MJ interaction energy on peptide $A_i$ at the interaction site $t^I_l$ caused by the sum of the interactions between $A_i$ and the protein residues in $P_l$ by $\tilde\epsilon_{il}$.

The goal of the optimization problem is to find the docking conformation for the peptide $A$ on the tetrahedral lattice $\mathcal{T}$ in the presence of the external protein $P$ that minimizes the aggregate MJ interaction potential while satisfying the cyclization conditions $C$ and ensuring peptide particles do not overlap with each other or with the set $\mathcal{T_B}$.
Additional details and assumptions of the problem definition can be found in Appendix \ref{appendix:re}. 

\subsection{QUBO Model}

Various QUBO formulations of the problem without the external protein interaction (i.e., without docking) with different encoding strategies have been previously introduced and analyzed. We give a detailed overview of the different approaches in Appendix~\ref{appendix:related_work}. Because of the beneficial scaling in the number of required variables, we start from the QUBO derived in~\cite{Robert_2021} and introduce two additional terms, one for peptide cyclization, and another for the influence of the external target protein. 

Our starting point is the Hamiltonian derived in~\cite{Robert_2021}, which solves the peptide folding problem without cyclization and protein interaction:
\begin{equation}\label{eq:tetra_H0_main}
H = H_{\rm comb} + H_{\rm back} 
\end{equation}
where $H_{\rm comb}$ combines the inter-peptide interaction energy with the no-overlap constraint penalty term for peptide particles that are more than three turns apart, and $H_{back}$ is the no-overlap constraint penalty term ensuring that subsequent main- and side-chain particles do not backtrack on each other. The Hamiltonian is based on a turn encoding of the peptide main- and side-chain particles, and the location of subsequent particles are defined by the turns on the tetrahedral lattice. For the explicit form of the Hamiltonian terms, we refer to Appendix \ref{appendix:re}. 

As our first extension of the model presented in~\cite{Robert_2021}, we add a cyclization constraint penalty term $H_{\rm cycle}$ that enforces the cyclization bond $(i,j) \in C$, between given peptide residue pairs $i$ and $j$. 
To do so, we define a distance function $d(X,Y)$ that calculates the squared number of steps along the lattice between two lattice vertices $X$ and $Y$, used to calculate distance between particles on the lattice. 
The distance is expressed in terms of the turn directions and the definition can be found in Appendix \ref{appendix:re}.
We enforce cyclization by constraining the particles of the residue $i$ and $j$ to be 1-NN on the lattice (i.e., $d(A_i,A_j)=1$).
For this we construct the term
\begin{equation}
    H_{\rm cycle} = \lambda_{\rm cycle} (d(A_i,A_j) - 1)^p \, ,
\end{equation}
where $\lambda_{\rm cycle}$ is a strictly positive scalar used to regulate the penalty strength and $p$ can be set to $1$ or $2$ depending on the solver properties.
$H_{\rm cycle}$ is equal to zero if the particles are one lattice step apart and adds a positive contribution otherwise as the non-overlap constraint forbids a distance zero. 

For the peptide-protein interaction, we add MJ interaction energies for peptide residues residing on lattice points within the interaction range of the protein $\mathcal{T_I}$, and a constraint penalty term to represent steric hindrance on the sites $\mathcal{T_B}$ blocked by the protein. 
We leverage knowledge of the problem geometry and assume that for any point in $\mathcal{T_B}$ all 1-NN points not in $\mathcal{T_B}$ are in $\mathcal{T_I}$. 
For this to hold it is sufficient to choose blocking and interaction radii that differ by at least one lattice step.
Necessarily, the peptide starts on a vertex outside of $\mathcal{T_B}$, and by our assumption, for any potential overlap with a point in $\mathcal{T_B}$ there must be an overlap with a point in $\mathcal{T_I}$.

Borrowing from the logic in $H_{\rm comb}$, we design a Hamiltonian that adds a protein-peptide MJ interaction energy when a peptide particle resides on a point in $\mathcal{T_I}$ and none of its bonded 1-NN main or side-chain residues are on a blocking point in $\mathcal{T_B}$, and is strictly positive otherwise. 
If any bonded 1-NN particles do land on a point in $\mathcal{T_B}$, they will incur a positive penalty to the system energy, driving the optimizer to turn off the particle contributions which then zeroes out the MJ interaction.
Additionally, we make sure that each interaction site can contribute at most one protein-peptide MJ interaction to the system, penalizing the placement of multiple peptide particles on the same interaction site. 

We introduce an interaction variable $\eta_{il}$ ($\eta_{i^sl}$) for any valid combination of main (or side) chain particle $i$ ($i^s$) and interaction site $t_l^{I} \in \mathcal{T_I}$, which acts as a switch in the minimization, turning on or off the energy contribution depending on if it is beneficial (negative contribution) or harmful (positive contribution) to the system.
By valid combination, we mean that the site $t_l^{I}$ has sufficient non-blocked 1-NN sites required by the bonds of the peptide particle. 
We then define the peptide-protein Hamiltonian by 
\begin{eqnarray}
    H_{\rm protein} & = & \sum_{i,l} \eta_{il}  \Big(\epsilon_{il}  (1 - \mu_1 d(A_i,t_l^{I})) + \mu_2 \sum_{X\in N(A_i)}\sum_{Y\in M_X(t_l^I)} \big(2 - d(X,Y)\big) \Big) \label{eq_Hprotein_1}\\
                &   & + \sum_{i,l} \eta_{i^sl} \Big(\epsilon_{il}  (1 - \mu_1 d(A_{i^s}^s,t_l^{I})) + \mu_2 \sum_{Y\in M_{A_i}(t_l^I)} \big(2 - d(A_i,Y)\big) \Big) \label{eq_Hprotein_2} \\
                & &  + \mu_3 \sum_l \Big( \sum_{i,j; i > j} (\eta_{il} \eta_{jl} + \eta_{i^sl} \eta_{j^sl}) + \sum_{i \geq j} \eta_{i^sl} \eta_{jl}  \Big)\, .\label{eq_Hprotein_3}
\end{eqnarray}
The outer sums run over all valid pairs of peptide main chain particle $i$ and lattice points $t_l^{I} \in \mathcal{T_I}$, indexed by $l$. 
$M_X(t^I_l)$ denotes the set of all lattice sites 1-NN of $t_l^I$ that are either in $\mathcal{T_B}$ or do not have sufficient non-blocked 1-NN for the peptide particle $X$ to be placed on them.
In the second sum on line~(\ref{eq_Hprotein_2}), the sum over all 1-NN of a side chain $A_{i}^s$ can be simplified as the only bonded particle is the corresponding main chain residue.
The sums on line~(\ref{eq_Hprotein_3}) are meant over all triples $l,i,j$ such that the corresponding pairs in the summands are valid, and thus for which interaction variables exist. 

The different terms in $H_{\rm protein}$ and a proof that the Hamiltonian satisfies the requirements discussed above can be found in Appendix~\ref{appendix:proofs}. 
Therein, we also derive the sufficient bounds for the penalty parameters $\mu_2 > -\epsilon_{il}/2$ for all $i,l$ and 
\begin{equation}
    \mu_1 \geq 1 + 9 \mu_2  |N(A_i)||N(t_l^{I})|/|\epsilon_{il}|, 
\end{equation}
where $|\cdot|$ applied to sets denotes the cardinality. Combining the Hamiltonian in Eq.~(\ref{eq:tetra_H0_main}) with the cyclization and protein interaction contributions, we arrive at our final Hamiltonian
\begin{equation}
    H = H_{\rm comb} + H_{back}  + H_{\rm cycle} + H_{\rm protein} \, .
\end{equation}

\subsection{CP Model}

For each residue $A_i$ we associate a vector of integer variables $(x_i, y_i, z_i) \in \mathbb{Z}^3$ to describe the position of the residue on the lattice. 
We use $s$ superscripts to identify side-chain elements where necessary (e.g., $x_i^s$ is the x-coordinate of the side-chain element for residue $A_i$). 

We also introduce integer variables that represent the pairwise dimension distance between residues: $(x^{\delta}_{ij}, y^{\delta}_{ij}, z^{\delta}_{ij}) \in \mathbb{Z}^3, \forall i \neq j$. 
We introduce a set of boolean variables that identify if a pair of residues (or a residue and an external protein residue) is within an interaction threshold $\psi$ on the lattice, and thus have a contribution to the objective function: $\alpha_{ij} \in \{0,1\}, \forall i \neq j$. 
Finally, we use a set of boolean variables that flag if a pair of residues have a different location along the axes: $(x^{\neq}_{i,j}, y^{\neq}_{i,j}, z^{\neq}_{i,j}) \in \{0,1\}^3, \forall i \neq j$.

In contrast to linear programming-based model-and-solve technologies, CP supports logical constraints and non-linear relationships. 
The core objective, constraints, and variables of our CP model with a standard energy-minimizing objective function are expressed by: 
\begin{align}
\min \quad & \sum_{i<j: j \geq i+2 \in A \times A} \alpha_{ij}\epsilon_{ij} + \sum_{i, j^s \in A \times A } \alpha_{i,j^s}\epsilon_{i,j^s}  \nonumber\\
& + \sum_{i^s, j^s \in A\times A} \alpha_{i^s,j^s}\epsilon_{i^s, j^s} + \sum_{i \in A, j \in P} \alpha_{i,j} \epsilon_{i,j} + \sum_{i^s \in A, j \in P} \alpha_{i^s,j} \epsilon_{i^s,j}  & \label{core-obj}  \\
\text{s.t.} \quad & \textsc{table}((x_i, y_i, z_i), \mathcal{T} \setminus \mathcal{T_B}) & \forall i, i^s \in A  \label{core-table}\\
& x^{\delta}_{ij} = |x_i - x_j|, y^{\delta}_{ij} = |y_i - y_j|, z^{\delta}_{ij} = |z_i - z_j| & \forall i \neq j \in A \times A \label{core-distance-linking} \\
& x^{\delta}_{i,i+1} + y^{\delta}_{i,i+1} + z^{\delta}_{i,i+1} = \Omega & \forall i \in A \setminus {A_N} \label{core-neighbor-distance} \\
& x^{\delta}_{i,i^s} + y^{\delta}_{i,i^s} + z^{\delta}_{i,i^s} = \Omega & \forall i \in A \label{core-neighbor-distance2} \\
& x^{\delta}_{i,j} + y^{\delta}_{i,j} + z^{\delta}_{i,j} = \Omega & \forall (i,j) \in C & \label{core-cyclization} \\ 
& x^{\neq}_{ij} \rightarrow x_i \neq x_j, y^{\neq}_{ij} \rightarrow y_i \neq y_j, z^{\neq}_{ij} \rightarrow z_i \neq z_j & \forall i\neq j \in A \times A \label{core-overlap-linking} \\
& x^{\neq}_{ij} \vee y^{\neq}_{ij} \vee z^{\neq}_{ij} & \forall i \neq j \in A\times A  \label{core-no-overlap} \\ 
& \alpha_{i,j} \rightarrow x^{\delta}_{i,j}, y^{\delta}_{i,j}, z^{\delta}_{i,j} \leq \psi & \forall i \neq j \in A \times A \label{core-interaction-linking} \\
& \alpha_{i,j} \rightarrow x^{\delta}_{i,j} + y^{\delta}_{i,j} + z^{\delta}_{i,j} \leq \psi_{total} & \forall i \neq j \in A \times A \label{core-interaction-linking2} \\ 
& \alpha_{i,j} \rightarrow (x_i, y_i, z_i) \in \mathcal{I}_j & \forall i \in A, j \in P \label{core-interaction-protein} \\ 
& (x_i, y_i, z_i) \in \mathbb{Z}^3_{+} & \forall i, i_s \in A  \label{core-var-start} \\ 
& (x^{\delta}_{ij}, y^{\delta}_{ij}, z^{\delta}_{ij}) \in \mathbb{Z}^{3}_{+} & \forall i \neq j \in A \times A \\ 
& (x^{\neq}_{ij}, y^{\neq}_{ij}, z^{\neq}_{ij}) \in \{0,1\}^3 & \forall i \neq j \in A \times A \\ 
& \alpha_{ij} \in \{0,1\} & \forall i \neq j \in A \times A \label{core-var-end} 
\end{align}
Objective \eqref{core-obj} expresses our core model objective function, which is to minimize MJ interaction energies of adjacent particles (since these interactions have negative values, minimization is desirable). 
This includes components for main-main, main-side, side-side, and external protein interactions. 
Constraint \eqref{core-table} uses the $\textsc{Table}$ constraint to ensure that the coordinates of each residue are placed on the points defined by the lattice but not affected by external protein steric hindrance, $\mathcal{T} \setminus \mathcal{T_B}$. 
The $\textsc{Table}$ constraint takes a tuple and maps it to a set of tuples (allowed value assignments). 
For example, if an external protein residue was located at coordinate $(1,1,1)$, we could exclude this coordinate from $\mathcal{T}$ to ensure the solver does not place a residue there. 

Constraint \eqref{core-distance-linking} links the pairwise distance variables to the location variables. 
Constraints \eqref{core-neighbor-distance}--\eqref{core-neighbor-distance2} link the summation of the pairwise distance variables (for both main chain and side-chain) in each dimension to be equal to $\Omega$, the rectilinear distance of adjacent neighbors in the lattice. 
Similarly, Constraint \eqref{core-cyclization} enforces neighboring rectilinear distance on residues included in cyclization bonds (for simplicity we omit constraints for cyclization of side-chain bonds, but these are handled in similar fashion). 
Constraint \eqref{core-overlap-linking} links the no-overlap variables and location variables, and Constraint \eqref{core-no-overlap} enforces the various non-overlap conditions for main-main, side-side, and main-side (namely that at least one coordinate value must be different between pairs). 
Constraints \eqref{core-interaction-linking}--\eqref{core-interaction-linking2} link interaction variables, $\alpha$, to pairwise distance variables and bound their domains according to the interaction threshold. 
Constraint \eqref{core-interaction-protein} enforces that a peptide residue must be located at one of the lattice vertices under the influence of the external protein (via projection), $\mathcal{I}_j$, for it to be considered interacting for the purposes of the objective function. 
Finally, Constraints \eqref{core-var-start} through \eqref{core-var-end} define the model variables and their domains. 

\subsection{Problem instances}\label{sec:instance-gen}

To measure the effectiveness of our approach, we take entries from the RCSB Protein Databank (PDB), and attempt to predict the peptide conformation found in each. 
By using PDB contents, we have ground truth solutions we can measure our predictions against.
A list of the PDBs we consider in this work is given in Table \ref{table:problem-instances}. 

\begin{table}[ht]
\centering
\begin{tabular}{cccccc}
\toprule
\multicolumn{1}{l}{\textbf{PDB ID}}  &  \textbf{Peptide Chain}  &  \textbf{Cyclization}  &  $|A|$  &  $|A^s|$  &  $|P|$  \\ \midrule

3AV9   &     S-A-K-I-D-N-L-D &     S1-D8 &     8 &     8 &     28 \\ 
3AVI   &     S-L-K-I-D-N-M-D &     S1-D8 &     8 &     8 &     32 \\ 
3AVN   &     S-H-K-I-D-N-L-D &     S1-D8 &     8 &     8 &     28 \\ 
3WNE   &     P-K-I-D-N-G     &     P1-G6 &     6 &     5 &     26 \\ 
5LSO   &     K-S-R-W-D-E     &     K1-E6 &     6 &     6 &     34 \\ 
2F58   &    H-I-G-P-G-R-A-F-G-G-G & H1-G10 & 11 & 6 & 49 \\
\bottomrule
\end{tabular}
\caption{RCSB Protein Databank (https://www.rcsb.org/) peptide-protein complexes used for our experimental analysis. Cyclization for peptide 5LSO is specified between side-chain particles, and the cyclization of remaining instances all involve main chain particles.
}
\label{table:problem-instances}
\end{table}

Instances in the PDB repository describe cyclic protein complexes using atomic-level coordinates in a standardized format. 
Before the PDB contents can be incorporated into the problem, they must be processed and registered onto the lattice. 
We provide a summary of the pipeline constructed to achieve this, but further details can be found in Appendix \ref{appendix:experiments}. The first step is to separate the information in the PDB file into two components: i) the peptide, and ii) the target protein (i.e., the protein that the peptide docks onto).
The next step is to identify the protein active site (see Figure \ref{fig:workflow_frame2}), which constitutes a smaller, more focused area that the peptide is likely to use for docking. 

Each atom in the peptide or protein active site is then classified as belonging to main chain or side chain, and the weighted centroid of each grouping is calculated, yielding a 2-particle representation for each amino acid in each structure.
The last step is to produce the tetrahedral lattice that discretizes the search space for our binary optimization approach (Figure \ref{fig:workflow_frame4}).
Our approach for placing the tetrahedral lattice is described in Algorithm \ref{alg:lattice-gen}. 
The lattice is then filtered according to the current active site coordinates $P$ and steric clash radius to remove vertices that would clash with the target site $\mathcal{T_B}$, yielding the final lattice structure. 
The problem instance can then be characterized by the peptide main and side chains, filtered tetrahedral lattice coordinates $\mathcal{T}'$ (with origin $t_1$), and shifted protein active site coordinates, $P$.

\section{Results}\label{sec:experiments}

We conduct an experimental analysis on the peptide instances summarized in Table \ref{table:problem-instances} using the instance generation pipeline and models discussed in Section \ref{sec:problem}. 

\subsection{QUBO constraint violations}

Because our resource-efficient turn encoding approach is a quadratic \emph{unconstrained} binary optimization formulation, there is the possibility that solutions correspond to peptide conformations that violate problem constraints. 
Further, as the problem size (number of  variables) grows, it becomes harder for the solver to find high-quality solutions that are feasible.
As such, we start by assessing the solver's ability to find feasible solutions for each problem instance. \\ 

\noindent{\bf Without external protein.}
By removing the external protein from the problem (and the associated penalty term $H_{protein}$), we reduce the QUBO problem size significantly, thereby making it easier to solve. 
For instance, for peptide 3WNE, the number of QUBO variables in the problem without the protein is 152, and the number of variables when the protein is included is 596.
This reduction in complexity leads to the contrasting outcomes between runs with and without the protein.
Without the protein, the model is able to find optimal solutions for each problem in the best case, though not in all cases.
Table \ref{table:constraint-violations-noprotein} summarizes the model's ability to find feasible solutions, and also reports on the best solution found per problem.
Further, while we did run hyper-parameter optimization (HPO) over these problems, we found that it was possible to use one general set of hyperparameters and yield feasible results. \\

\begin{table}[h!]
\centering
\begin{tabular}{ccc}
\toprule
\textbf{PDB ID} &  \textbf{Feasible Fraction} & \textbf{Best MJ Interactions} \\
\midrule
3AV9 &  0.55 & -6.64 \\
3AVI &  0.61 & -8.54 \\
3AVN &  0.55 & -6.88 \\
3WNE &  1.0  & -1.87 \\
5LSO &  1.0  & -5.97 \\
2F58 &  0.08 & -6.02 \\
\bottomrule
\end{tabular}
\caption{QUBO solutions with \textbf{no protein} present. The table shows the fraction of feasible solution and the lowest MJ interaction across all 100 shots for each problem instance.
}
\label{table:constraint-violations-noprotein}
\end{table}

\noindent{\bf With external protein.}
Here we compare against runs where the external protein was included.
The QUBO model was not able to solve all problem instances in this case, so we switch the analysis to instead report on the number and type of violations observed.
For each problem instance, we first determine the shot that resulted in the most optimized QUBO objective function value and then assess the breakdown of constraint violations for that solution. 
We report violations for the number of overlaps involving pairs of peptide residues (\# Overlaps), the number of steric clashes between a peptide residue and the external protein (\# Protein Clash, for runs that included the protein) and whether the cyclization constraint is satisfied in the solution.

As shown in Table \ref{table:constraint-violations-protein}, the QUBO approach yields feasible solutions for smaller peptides 3WNE and 5LSO, but struggles as problem size increases.
For instance, the best MJ solution for 3AV9 has six overlaps of the peptide, two steric clashes with the external protein, and the cyclization was not satisfied. 
This is not a practically useful solution, but it is worth noting that from the QUBO perspective, it is not a particularly bad solution.
Because 3AV9 involves eight main chain residues and eight side chain residues (for a total of sixteen peptide residues), there are ${\tiny {16 \choose 2}} = 120$ possible pairs of overlapping residues. 
This means that the QUBO-based solution satisfies $114/120$ $(95.0\%)$ of the peptide overlap constraints.
When considering the external protein, the 3AV9 instance involves 28 external protein residues indicating a possible $16 \times 28 = 448$ steric clash combinations. 
Of these, the solution violates 2, satisfying over 99.5\% of these constraints.
Note that this calculation provides a lower bound, and is useful for illustrative purposes; the projection of blocking influence onto the lattice often yields more blocking sites on the lattice than the number of residues present, and the peptide could potentially clash with any of those.
 
\begin{table}[h!]
\centering
\begin{tabular}{cccc}
\toprule
\textbf{PDB ID}  &  \textbf{\# Overlaps (\% Sat.)}  &  \textbf{\# Protein Clash (\% Sat.)}  &  \textbf{Cyclization} \\
\midrule
3AV9   &     6 (95.0\%) &      2 (99.6\%) &     False \\
3AVI   &     9 (92.5\%) &      1 (99.8\%) &      True \\
3AVN   &     5 (95.8\%) &      1 (99.8\%) &     False \\
3WNE   &     0 (100.0\%) &     0 (100.0\%) &      True \\
5LSO   &     0 (100.0\%) &     0 (100.0\%) &      True \\
2F58   &     14 (89.7\%)   &      2 (99.8\%)    &     True   \\
\bottomrule
\end{tabular}
\caption{\textbf{With protein} QUBO solution constraint violations by instance. For each peptide, the shot associated with the best MJ interaction is selected (out of 10 shots total). Overlaps and protein clash calculated for all relevant pairs $(i <j)$. \% Satisfied (\% Satis.) calculated as the number of conflicts observed divided by the total number of possible conflicts. Peptide residue count (\# Peptide) includes main and side chain residues.
}
\label{table:constraint-violations-protein}
\end{table}

Our results demonstrate that mapping protein optimization type problems to QUBOs inherently makes these harder to solve, not only because penalty terms are hard to construct and parameters hard to tune, but also because any solver must find sufficiently high-quality (e.g., low energy) solutions to avoid the violation of hard constraints, which is not always possible, especially for such large systems. This also means that protein-related problems are not well suited for QUBO-based solvers, e.g., quantum annealers \cite{hauke:20}.

\subsection{MJ interactions and RMSD}

Here we assess the solutions produced by the QUBO approach in terms of both their MJ interaction value and the root mean square distance (RMSD) when compared with the coordinates of the real peptide (sourced from the PDB file) in Euclidean space. 
The RMSD measures the average distance between the coarse-grained particles of the real (PDB) peptide and the peptide produced by our approach. 
The equation for RMSD is defined as follows:
\begin{equation}
    \mathrm{RMSD}(\mathbf{v}, \mathbf{w}) = \sqrt{\frac{1}{n}\sum_{i=1}^{n} ||v_i - w_i||^2} ,
\end{equation}
where $\mathbf{v}$ denotes the coordinates for the real peptide and $\mathbf{w}$ denotes the coordinates of the QUBO-generated peptide,
and the RMSD represents the $\ell^2$-norm of the difference between both vectors. 

It is important to note that while our approach optimizes for aggregate MJ energy, the practical success of a conformation is typically gauged using RMSD or other molecular similarity metrics. 

The results in Tables \ref{table:mj_rmsd_results-noprotein} and \ref{table:mj_rmsd_results-protein} detail, for the best solution for each peptide instance, the associated values of the objective function (QUBO MJ) and RMSD to the real peptide. 
To provide context for these results, we also include the optimal MJ and associated RMSD attained by our CP-based approach, where the CP model is optimizing aggregate MJ interaction energy as in Objective \eqref{core-obj}, with the first residue anchored to the origin to facilitate fair comparison with the QUBO results. \\

\noindent {\bf Without external protein.}
Removing the external protein removes the external (protein) MJ interaction potentials associated with the ``correct'' conformation; meaning there is no force driving the solver to a conformation that is spatially aligned with the ground truth PDB result, only those resulting from a feasible cyclic conformation.
This also means that the peptide could fold into a stable, feasible conformation in a number of directions that it otherwise would be driven away from, thereby making it easier to find a feasible conformation.
Removal of the protein also removes the protein blocking influence over the peptide, meaning a feasible solution in this context could actually be infeasible if the external protein were considered.

\begin{table}[h]
\centering
\begin{tabular}{cccccccc}
\toprule
\textbf{PDB ID} &  MJ: \textbf{QUBO} &  \textbf{CP (1-NN)} &  \textbf{CP}  &  RMSD (\AA): \textbf{QUBO} &  \textbf{CP (1-NN)} &  \textbf{CP} &  \bm{$t_{\rm QUBO}$}\textbf{(s)} \\
\midrule
3AV9  &     -6.64 &     -6.64 &     -101.09 &      11.31 &     11.70 &    10.81 &     153 \\
3AVI  &     -8.54 &     -8.54 &     -114.71 &       9.14 &      9.36 &     8.87 &     197 \\
3AVN  &     -6.88 &     -6.88 &     -102.00 &      10.52 &     11.32 &    12.55 &     202 \\
3WNE  &     -1.87 &     -1.87 &      -46.30 &       9.76 &      8.17 &     7.51 &     122 \\
5LSO  &     -5.97 &     -5.97 &      -53.56 &      13.32 &     13.64 &    13.64 &     161 \\
2F58 &  -6.02 & -13.61 & -134.19 & 11.06 & 9.85 & 9.85 & 215 \\
\bottomrule
\end{tabular}
\caption{\textbf{No protein} MJ interaction and RMSD values for QUBO approach vs. CP approach. The CP model is anchored to the same starting point as QUBO method to enable fair comparison. The CP model includes runs with interaction radius limits set to one grid step (1-NN) and the default interaction radius (6.5\AA).}
\label{table:mj_rmsd_results-noprotein}
\end{table}

We note an obvious gap between the QUBO and CP MJ energies using the default interaction radius of 6.5\AA.
When the CP model uses a reduced interaction radius (one grid step), we see close alignment between the QUBO and CP (1-NN) results, showing that the QUBO model is able to effectively solve the problems.
We also see that the runtimes are fairly consistent across peptides, with 3WNE being the fastest at 122s ($\sim$2min), and 3AVN being the slowest at 202s ($\sim$3.5min).
This leads to the expectation that the solver can solve problems without the external protein within $\sim$5min. \\

\noindent{\bf With external protein.}
Now we include the external protein and run the same experiments.
Table \ref{table:mj_rmsd_results-protein} summarizes our results.
First, the QUBO model yields worse MJ energies than the equivalent CP (1-NN) model, and is expectantly worse than the default CP model.
The 3AV* problem instances show high absolute aggregate MJ values, but this is due to infeasible conformations that lead to additional interactions with external protein residues that are impractical.
Second, where the QUBO model yielded feasible solutions (3WNE, 5LSO), it produced results that were closer to the PDB solution per the RMSD metric.

\begin{table}[ht!]
\centering
\begin{tabular}{cccccccc}
\toprule
\textbf{PDB ID} & MJ: \textbf{QUBO} &  \textbf{CP (1-NN)} &  \textbf{CP}  & RMSD (\AA): \textbf{QUBO} &  \textbf{CP (1-NN)} &  \textbf{CP}  &  \bm{$t_{\rm QUBO}$}\textbf{(s)} \\
\midrule
3AV9   &    -115.86* &     -46.56 &     -133.74 &    7.71* &     9.93 &     8.98 &     658 \\
3AVI   &    -169.17* &     -62.34 &     -159.81 &    8.10* &     9.85 &     8.72 &     942 \\
3AVN   &    -129.67* &     -55.12 &     -136.83 &   10.55* &     8.45 &     8.23 &     657 \\
3WNE   &      -25.19 &     -42.64 &      -76.86 &     7.45 &     8.48 &     8.48 &     296 \\
5LSO   &      -17.34 &     -23.26 &      -62.41 &     7.68 &    10.29 &    10.29 &     239 \\
2F58 &  -260.53*  &  -4.02 & -215.91  & 7.55 & 12.05 & 10.42 & 1560 \\
\bottomrule
\end{tabular}
\caption{\textbf{With protein} MJ interaction and RMSD values for QUBO approach vs. CP approach. The CP is model anchored to the same starting point as QUBO method to enable fair comparison. Entries with asterisk (*) denote infeasible results (violating problem constraints).}
\label{table:mj_rmsd_results-protein}
\end{table}

\begin{figure}[ht!]
\centering
\includegraphics[width=0.9\textwidth]{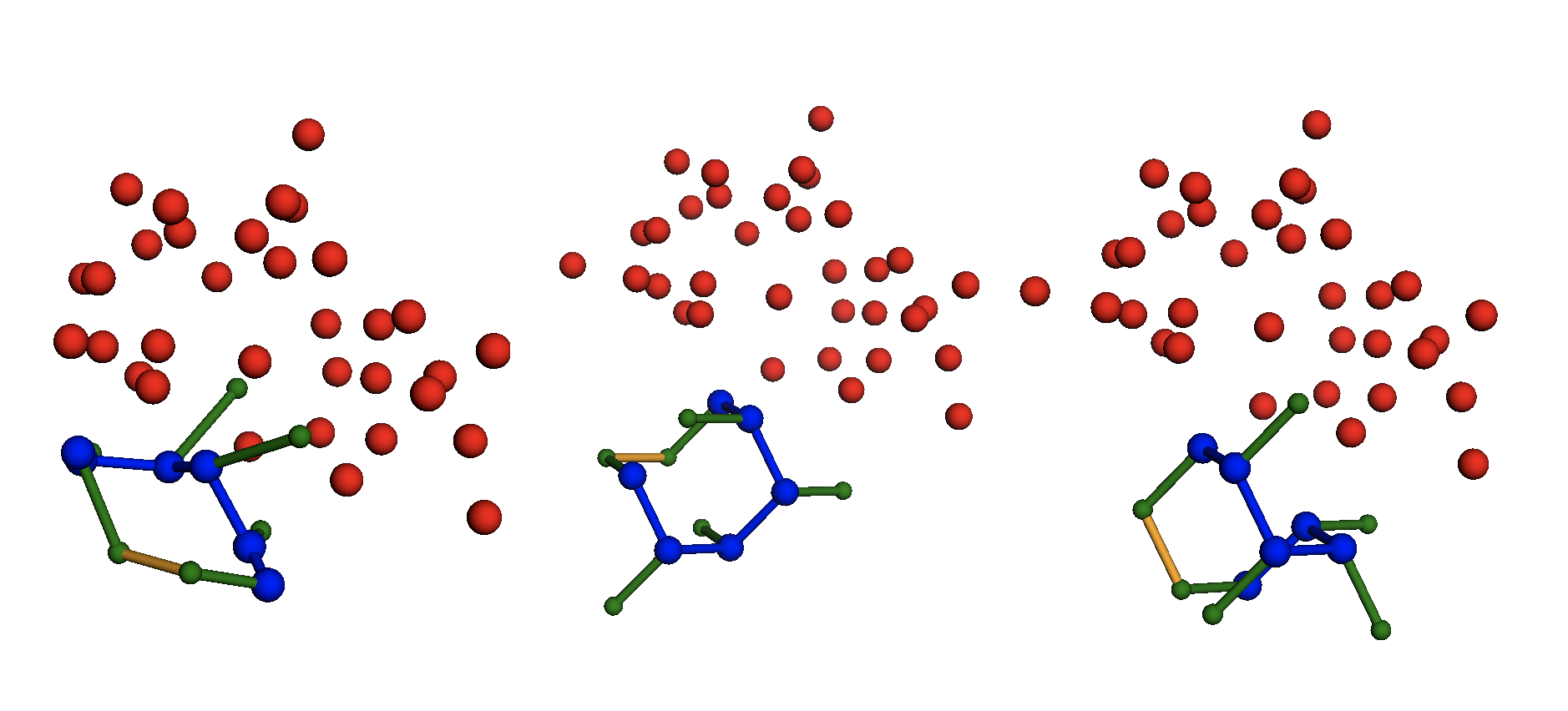}
\caption{5LSO result visualization. Comparing the true PDB peptide (left) against the QUBO model result (middle) and the CP-MJ 1-NN result (right). The peptide is represented with blue and green dots (main and side-chain, respectively), and the red dots are the external protein residues.}
\label{fig:5lso-result-comparison-plot} 
\end{figure}

This may seem surprising, but we must consider that the CP results shown here are for runs of the CP model where the first peptide residue is \textbf{anchored} to the lattice origin.
Recall that the anchor was introduced in order to provide a more direct and fair comparison between QUBO and CP models.
Because of this anchoring, while working to maximize interactions with the external protein, the CP models create high-quality solutions that are rotated off of alignment with the ground truth PDB result.
An example of this is shown in Fig.~\ref{fig:5lso-result-comparison-plot}.
Finally, we note some significant variance in solver runtimes in this case, with 5LSO being the fastest at 239s ($\sim$4min), and 3AVI being the slowest at 942s ($\sim$16min).

\section{Discussion}\label{sec:conclusions} 

In this work, we have explored solving the cyclic peptide docking problem on a fixed lattice in a general manner using a QUBO formulation, and in parallel constructed a novel constraint programming (CP) formulation.
To do this, we have built on the work of Ref.~\cite{Robert_2021} to include a cyclization constraint term and a general peptide-protein interaction term, which had not been done before.
We find that wrapping the combined distance checks on interaction and blocking sites with an ancilla bit (to turn on/off), as described in Sec.~\ref{sec:problem}, is the most efficient approach to include the protein interactions.

One of the primary difficulties of modeling cyclic peptide docking problem in the QUBO formalism is the construction of disjunctive expressions that turn `on' or `off' various components of the function.
The problem requires inclusion or exclusion of forces (energies) according to distance thresholds (e.g., overlap penalties when $d_{ij}=0$ or pairwise MJ interactions when $d_{ij}=1$), which exist at single points in space and are zero otherwise. 
These would be effectively modeled by a (set of) delta function(s).
Unfortunately, there is no easy or efficient way to include these types of forces naturally in a Hamiltonian expression.
Instead, we must add ancilla bits which leads to additional overhead in the problem representation and makes it harder to solve.

An important simplification to the cyclic peptide docking problem is the restriction of placement of the residues on vertices of a lattice structure.
In practice, peptides are not constrained to a particular rigid underlying structure, they simply abide by steric forces preventing the molecules from clashing with one another.
Therefore, the introduction of any lattice structure introduces some error between the model result and the true PDB solution. 
This is one driver of the large RMSD values seen in Tables \ref{table:mj_rmsd_results-noprotein} and \ref{table:mj_rmsd_results-protein}.

When the external protein is removed, all peptides tested can be solved with the QUBO representation. 
As shown in Table \ref{table:mj_rmsd_results-noprotein}, the solution quality of these ``no protein'' simulations were quite high, with QUBO model MJ potentials equal to those found by the CP model in four out of five cases with 3AVN being the exception where the QUBO model yielded a total MJ potential of $-6.63$ and the CP model yielded $-6.88$ (approx.~3.63\% gap).
Because the protein is absent in these simulations, we omit the calculation of the RMSD comparison to the PDB solution.

When the external protein is included, the QUBO approach finds feasibility for only two out of five peptides (3WNE, 5LSO).
Within these, as Table \ref{table:mj_rmsd_results-protein} shows, the  MJ potentials were both worse than the CP solutions, with a gap of approximately 40.92\% for 3WNE and approximately 25.45\% for 5LSO.
The RMSD between the feasible QUBO results and the PDB results (average 7.57\AA) is better than the RMSD between the CP results and the PDB results for the same peptides (average 9.39\AA). This suggests that MJ-optimized solutions may not always perform well with respect to RMSD against the PDB ground truth, especially when anchored to what might be a sub-optimal starting location.
Further, it is worth noting that in practice, an RMSD greater than 4\AA \ is considered to be a non-match and not practically useful. 

Overall, the inability of the QUBO model to consistently find feasible solutions in the presence of an external protein is a significant limitation of the approach, which also implies that quantum optimization techniques \cite{hauke:20} that leverage QUBO representations might not be the most efficient for this class of application. In contrast, the CP approach is scalable and has the potential to deliver new insights into peptide-protein interactions. 

\bibliographystyle{elsarticle-num}
\bibliography{paper}

\section{Acknowledgements}\label{sec:ack} 
\subsection{Acknolwedgements}
The authors would like to thank Grant Salton, Gili Rosenberg, Yusuke Mastumoto, and Shoko Ustunomiya for their support in this work through their rigorous feedback and various insightful discussions. The authors would also like to thank Takuya Shiraishi and Yoshihisa Murata for engaging in productive research discussions and providing suggestions, and Yuji Koriyama and Masako Kimura for their support on contractual procedures. 

\subsection{Data availability}
We source all our raw data from the RCSB Protein Databank (PDB), which can be found at https://www.rcsb.org/ .
Due to IP concerns, we do not provide exports of our processed PDB intermediates, but we do provide a description of the steps necessary to recreate the data in Appendix \ref{appendix:experiments}

\subsection{Author contributions}
Members of the Amazon team, including Kyle Brubaker, Kyle Booth, Fabian Furrer, and Jayeeta Ghosh developed the pipeline code, ran the experiments, and conducted the data analysis for the experiments.
Helmut Katzgraber assisted in the conceptualization of the various QUBO approaches and in the experimental design.
Members of the Chugai team including Akihiko Arakawa and Tsutomu Sato supported in conceptualization of the approach and experimental design, provided supporting code implementation, and provided regular feedback and subject matter expertise.

\subsection{Additional Information}

\textbf{Competing Interests} The authors declare no competing interests, financial or otherwise.

\begin{appendices}

\section{Related work} \label{appendix:related_work}

Perdomo et al.~\cite{Perdomo_2008} solved for the optimal conformation of an amino acid sequence on a two-dimensional grid, using the hydrophobic-polar (HP) interaction model and a one-particle representation leveraging quantum computing techniques.
This work introduced the idea of a spatial encoding, wherein the coordinates on the grid are mapped to sets of bits, and the goal is to assign bits to each residue such that their placement on the grid minimizes the system energy (thereby maximizing the number of hydrophobic (H) residues that interact).
While this spatial encoding might be the most intuitive encoding---it is straightforward to map a solution bitstring back to a series of $x$ and $y$ coordinates---it not very scalable for larger peptides in three-dimensional lattice geometries because of the bit requirements to encode the position in each dimension \cite{Babej_2018}. 

An alternative to spatial encoding is turn encoding, which defines the peptide geometry by a series of successive turns connecting one particle with the next, or to its respective side chain particle, see Ref.~\cite{Babbush_2014}. 
In the turn encoding there are a fixed number of turn directions, according to the lattice geometry, and each turn direction is mapped to a set of bits.
The solution bitstring defines the full series of turns the peptide particles take, starting from the first residue, to yield the final conformation.
This encoding offers an improved variable scaling proportional to the peptide length \cite{Babej_2018} relative to the spatial encoding.
Perdomo {\em et al}.~successfully applied this turn encoding to a small peptide of six amino acids, formulated as a one-particle coarse grained (CG) representation on a square lattice using a quantum annealer developed by D-Wave Systems Inc.~\cite{Perdomo_2012, DWave}.

Babbush {\em et al}.~\cite{Babbush_2014} provide an overview of multiple problem formulations based on turn encodings, and introducing an approach using slack variables to implement the overlap constraint as a penalty term, referred to as the  ``turn ancilla'' approach.
Babej {\em et al}.~\cite{Babej_2018} applied the turn ancilla approach to the cubic lattice with a dense turn encoding and a one-particle CG representation, to fold a 10-residue protein on a quantum annealer. 
Fingerhuth {\em et al.}~\cite{Fingerhuth_2018} extended this approach, using a one-hot encoding for each turn direction, which yielded a sparser representation allowing for alternative calculations of the Hamiltonian components. 
This was important for formulating the problem for a quantum alternating operator ansatz (QAOA) approach, which could be run on the Rigetti Aspen quantum circuit device \cite{Fingerhuth_2018, Rigetti}.
Robert {\em et al}.~\cite{Robert_2021} applied the turn encoding to model two-particle CG residues on a three-dimensional tetrahedral lattice, using a more variable efficient method to implement the overlap constraint than the turn ancilla approach.
They then applied a Variational Quantum Eigensolver (VQE) algorithm on the IBM Q 20-qubit circuit-based quantum device to find minimal energy states. 
Due to the limited number of qubits available, they reduced the problem scale in their experiments to a peptide consisting of seven residues, using the one-particle CG representation.
Boulebnane {\em et al}.~\cite{Boulebnane_2022} used a heavy atom representation of amino acids, wherein hydrogen atoms are grouped together with larger atoms such as carbon, oxygen, or nitrogen, placed on a tetrahedral geometry, modeling peptide interactions via the Lennard-Jones interaction potential. 
Furthermore, they introduced a relative turn encoding that maps the three possible directions into a qutrit and applied QAOA to solve to minimization problem.
This was a more granular problem representation, and therefore a more realistic approach, but the scaling requirement was high, even for a small four-residue peptide.

There has been limited research in formulating lattice-based models for protein-peptide docking experiments that could be run on quantum computers.
Perdomo {\em et al}.~\cite{Perdomo_2012} introduced some related ideas on accounting for chaperone molecules that impact the resulting peptide conformations.
This can be seen in their Supplemental Section I.C, where they assume the placement of the first two residues, and solve for the remaining length of four, accounting for a hypothetical molecule that blocks multiple points on the grid.
By knowing their starting point, they can construct a set of equations that would penalize solutions that would lead into this region or would incur a backtrack.
The work enumerates all such cases and constructs a composite formula to handle them. 
However, this is an intractable approach and is not applicable to arbitrary peptide and protein complexes.

Our work represents an extension of the problem formulation to the protein-peptide docking problem that applies to {\em arbitrary peptides and proteins}.
For a concise overview of the related work cited we refer to Table~\ref{table:lit-rev}, which summarizes problem setup and contribution.

\begin{table}[hbt!]
\centering
\begin{tabular}{|l|l|l|l|l|l|}
\hline
\textbf{Reference}  & \textbf{Year} & \textbf{Encoding}                                                                                                    & \textbf{Lattice}                                                        & \textbf{CG model}    & \textbf{Interaction} \\ \hline
Perdomo-Ortiz {\em et al.}~\cite{Perdomo_2008}       & 2008          & Spatial encoding                                                                                                     & 2D Square                                                               & 1-particle           & \begin{tabular}[c]{@{}l@{}}HP \\ Intra.\end{tabular}                   \\ \hline
Babbush {\em et al.}~\cite{Babbush_2014}      & 2012          & \begin{tabular}[c]{@{}l@{}}Turn ancilla, \\ turn circuit, \\ diamond\end{tabular}                                    & 2D Square                                                               & 1-particle           & \begin{tabular}[c]{@{}l@{}}HP \\ Intra.\end{tabular}                   \\ \hline
Perdomo-Ortiz {\em et al.}~\cite{Perdomo_2012}& 2012          & Turn ancilla                                                                                                         & 2D Square                                                               & 1-particle           & \begin{tabular}[c]{@{}l@{}}MJ \\ Intra.\end{tabular}                   \\ \hline
Fingerhuth {\em et al.}~\cite{Fingerhuth_2018}   & 2018          & \begin{tabular}[c]{@{}l@{}}One hot \\ turn encoding \\ QAOA\end{tabular}                                             & 2D Square                                                               & 1-particle           & \begin{tabular}[c]{@{}l@{}}HP and MJ \\ Intra.\end{tabular}            \\ \hline
Babej {\em et al.}~\cite{Babej_2018}        & 2018          & \begin{tabular}[c]{@{}l@{}}Turn ancilla, \\ also introduced \\ turn circuit and \\ spatial nested shell\end{tabular} & 3D Cubic                                                                & 1-particle           & \begin{tabular}[c]{@{}l@{}}MJ \\ Intra.\end{tabular}                   \\ \hline
Robert {\em et al.}~\cite{Robert_2021}      & 2021          & \begin{tabular}[c]{@{}l@{}}Turn encoding \\ (sparse and dense) \\ QAOA\end{tabular}                                  & 3D Tetrahedral                                                          & 2-particle & \begin{tabular}[c]{@{}l@{}}MJ \\ Intra.\end{tabular}                   \\ \hline
Irback {\em et al.}~\cite{Irback_2022}      & 2022          & Spatial encoding                                                                                                     & 2D Square                                                               & 1-particle           & \begin{tabular}[c]{@{}l@{}}HP \\ Intra.\end{tabular}                   \\ \hline
Boulebnane {\em et al.}~\cite{Boulebnane_2022}   & 2022          & \begin{tabular}[c]{@{}l@{}}Relative turn \\ encoding\end{tabular}                                                    & \begin{tabular}[c]{@{}l@{}}2D Square and \\ 3D Tetrahedral\end{tabular} & heavy atomic         & \begin{tabular}[c]{@{}l@{}}LJ \\ Intra.\end{tabular}        \\ \hline
\begin{tabular}[c]{@{}l@{}}{\bf Our work:} \\ Cyclic peptide docking\end{tabular}   & 2024          & \begin{tabular}[c]{@{}l@{}}Turn encoding\end{tabular}                                                    & \begin{tabular}[c]{@{}l@{}} 3D Tetrahedral\end{tabular} & 2-particle  & \begin{tabular}[c]{@{}l@{}}MJ \\ Intra. \\ + Inter.\end{tabular}   \\ \hline
\end{tabular}
\caption{\label{table:lit-rev}Literature review of related work. Here ``Intra.''~refers to intramolecular interactions and ``Inter.''~refers to intermolecular interactions. 2D refers to a two-dimensional square lattice and 3D to a three-dimensional cubic one, respectively.}
\end{table}
\label{table:related_work}

\section{QUBO formulation using resource-efficient turn encoding}\label{appendix:re}

Here we provide the building blocks of the resource efficient QUBO formulation by more thoroughly defining the elements of $H_{\rm comb}$, as initially described in Ref.~\cite{Robert_2021}. 

\subsection{Turn encoding on a tetrahedral lattice}

At each vertex there are four different turns that connect to the neighboring vertices (recall Figure \ref{fig:3dtetrahedral}).
Because of the geometry of the tetrahedral lattice, the direction of these turns alternate with each step taken on the lattice. 
Therefore, we introduce alternating sub-lattices $\mathcal{T_{\alpha}}$ and $\mathcal{T_{\beta}}$ with counter-directional turn directions.
On vertices of $\mathcal{T_{\alpha}}$, we chose an arbitrary but fixed enumeration of the four possible turns $(0,1,2,3)$. 
We then observe that on vertices $\mathcal{T_{\beta}}$ the turn directions are inverted and label them so that turn $k\in (0,1,2,3)$ on $\mathcal{T_{\beta}}$ is the inverse of $k$ on $\mathcal{T_{\alpha}}$. 
Hence, if turn $k$ points in the direction of vector $e_k$ on $\mathcal{T_{\alpha}}$ then turn $k$ on $\mathcal{T_{\beta}}$ points in the inverted direction $(-e_k)$.
On vertices of $\mathcal{T_A}$, we denote the four turn directions by $t \in (0,1,2,3) $ and on vertices $\mathcal{T_B}$ by $t \in (\bar{0},\bar{1},\bar{2},\bar{3})$, where turn direction $k$ and $\bar{k}$ are in opposite directions. 

Because the tetrahedral lattice has four turn directions, they can be bijectively mapped to pairs of bits via the encoding $ (0,1,2,3) \mapsto (00,01,10,11)$.
The conformation of the main-chain particles of a peptide of length $N$ is characterized by $N-1$ turns, and thus needs $2(N-1)$ bits to be encoded $(q_1q_2...q_{2N-3}q_{2N-2})$. 
Note that for the following formulas we assume that the main chain of the peptide starts on sub-lattice $\mathcal{T_{\alpha}}$ and the number of turns uniquely specifies the sub-lattice a particle is located on.

To encode the location of the side-chain particles, we use the same turn encoding applied to the turn $s_i$ connecting main- and side-chain particles $A_i$ and $A_i^s$.
Hence, the side chain is represented by $2N^s$ bits $\tilde{q}_1\tilde{q}_2....\tilde{q}_{2N^s-1}\tilde{q}_{2N^s})$, where $N^s \leq N$ is the number of side chain particles. 

In order to calculate MJ interactions and enforce constraints, we need to express the distance between any lattice vertices in terms of our turn encoding.
We will again follow Ref.~\cite{Robert_2021} and use turn indicators $f_k$, defined by $f_k(t) = 1$ if and only if $t$ is a turn in direction $k$ and otherwise $0$. 
These functions can be easily expressed in terms of the bit values $q_1,q_2$ of the turn $t$ by $f_0(t) = (1-q_1)(1-q_2)$, $f_1(t) = q_2(q_2-q_1)$, $f_2(t) = q_1(q_1-q_2)$ and $f_3(t) = q_1q_2$. 

Fixing the lattice origin at the start of the peptide, any lattice vertex $v$ can be identified by a four-dimensional turn vector $x = (x_0,x_1,x_2,x_3)$, where $x_k$ denotes the number of required turns in direction $k$ from the origin in order to reach $v$. 
We keep track of the inverted directions on different sub-lattices by setting $x_k = 1$ for a turn in direction $k$ on sub-lattice $\mathcal{T_{\alpha}}$, and $x_k = -1$ for a turn in direction $\bar{k}$ on sub-lattice $\mathcal{T_{\beta}}$. 
For any $A_i$, we can express $x_k$ using the turn indicators
\begin{equation}\label{eq:turn_representation}
x_k(A_i) = \sum_{l=1}^{i-1} (-1)^{l+1} f_k(t_l) 
\end{equation}
and for a corresponding side chain by
\begin{equation}
x_k(A_i^s)  = x_k(A_i) + (-1)^{i+1} f_k(s_i) \, .
\end{equation}
Using the turn vectors, we introduce the squared distance between any $A_i$ and $A_j$ by 
\begin{equation} \label{eq:tetra_dist}
d(A_i,A_j) = \sum_k\big( x_k(A_j) - x_k(A_i) \big)^2 \, .
\end{equation}
Note that the squared distance is used to ensure the distance value is non-negative, but comes at the cost of introducing higher-order polynomial terms into the Hamiltonian, which leads to variable overhead in the reduction of the problem to a QUBO.

This distance function can also be applied to arbitrary lattice points $v$ and $w$ with turn vectors $x$ and $y$ as $d(v,w) = \sum_k(y_k - x_k)^2$. 
An important property of the squared distance, resulting from the geometry of the tetrahedral lattice, is that the distance between any pair of grid points that are two turns apart is equal to $2$ and cannot be $4$ because two consecutive steps in the same direction lead to $0$ (recall $x_k = 1$ on $\mathcal{T_A}$ and $x_k = -1$ on $\mathcal{T_B}$). 

\subsection{Hamiltonian construction}

The Hamiltonian describing the optimization problem contains two categories of terms: the actual energy terms to be minimized, and the constrained penalty terms enforcing the validity of the solutions. 
The energy terms account for the interaction energies of non-bonded peptide particles and the peptide-protein interaction energy. 
The constrained penalty terms are constructed such that they are equal to zero if the constraints are satisfied and strictly positive otherwise.

Our starting point is the Hamiltonian derived in Ref.~\cite{Robert_2021}, which solves the peptide folding problem without cyclization and peptide-protein interaction. 
The Hamiltonian is given by 
\begin{equation}\label{eq:tetra_H0}
H = H_{\rm comb} + H_{\rm back} 
\end{equation}
where $H_{\rm comb}$ combines the inter-peptide interaction energy with the no-overlap constraint penalty term for peptide particles that are more than three turns apart, and $H_{\rm back}$ is the no-overlap constraint penalty term to avoid backtracking of subsequent main and side chain turns. 
For completeness, and because the formulas including side chains are not explicitly given in Ref.~\cite{Robert_2021}, we  provide them here. 

\subsubsection{Backtracking constraint}

We introduce the helper function $T(t,t') = \sum_{k=0}^3 f_k(t)f_k(t')$ that is equal to $1$ if turns $t$ and $t'$ are in the same direction and else $0$.
We can then formulate the no backtracking condition as
\begin{equation}\label{eq:Hback}
H_{\rm back} = \lambda_{\rm back} \Big( \sum_{i=1}^{N-2} T(t_i,t_{i+1}) + \sum_{i=1}^{N-1} T(t_{i},s_{i+1}) + \sum_{i=1}^{N-1} T(t_{i},s_{i}) \Big) \, ,
\end{equation}
where the first sum adds a positive penalty for backtracking along main-chain turns, the second for backtracking of side-chain turns on the main-chain turns, and the third for side-chain turns in the same direction as main-chain turns. 
Here we assume that every main-chain particle has a side-chain particle. otherwise the terms containing $s_i$ are omitted.
The constraint penalty parameter $\lambda_{\rm back} > 0$ can be adjusted and determines the Hamiltonian contribution of a constraint violation due to backtracking.

\subsubsection{Inter-peptide interaction energy combined with no-overlap constraint}

The inter-peptide interaction energies are added for all non-bonded particle pairs that are nearest neighbors (1-NN), i.e., one turn apart. 
The key idea behind the no-overlap constraint from Ref.~\cite{Robert_2021} is to exploit the fact that whenever there is an overlap between two amino acids, there must also be an inter-peptide interactions from its bonded nearest neighbors. 
For any non-bonded pairs $i,j$ a term is introduced that is equal to the MJ interaction energy $\epsilon_{ij}$ if a pair $i,j$ is one edge apart and the pair's bonded particles do not violate an overlap constraint, and is strictly positive otherwise.
This term will be multiplied with a binary interaction variable $\omega_{ij} \in \{0,1\}$, which is acting as a switch in the minimization, i.e, set to $0$ if the term is positive and $1$ if negative.
In the energy minimization search, this will prefer the conformations without overlaps, as only interaction energy is counted in those cases.

To properly account for main- and side-chain interactions, the combined Hamiltonian is decomposed into three terms 
\begin{equation}
    H_{\rm comb} = H_{\rm mm} + H_{\rm ms} + H_{\rm ss} \, ,  
\end{equation}
where $H_{\rm mm}$ contains the terms associated with a main chain-main chain (main-main), $H_{\rm ms}$ main-side and $H_{\rm ss}$ side-side interaction pairs, respectively.

We first consider the main-main interaction term and note that because of the tetrahedral geometry, two main-chain particles $A_i$ and $A_j$ can only be 1-NN if they are an odd number of turns apart and $|j-i| \geq 5$. The Hamiltonian in Ref.~\cite{Robert_2021} is given by
\begin{equation}\label{eq:H_mm}
    H_{\rm mm} = \sum_{j\geq i+5\,, (j-i) \text{ odd}} \omega_{ij} ( \epsilon_{ij} + \lambda_1 (d(A_i,A_j)-1) + \lambda_2 K_{ij})  \, , 
\end{equation}
where $\lambda_1> 0$ and $\lambda_2 > 0$ are penalty parameters, $K_{ij}$ are the constraint penalty terms for the overlap checks on $A_i$ with the 1-NN of $A_j$ and $A_j$ with the 1-NN of $A_i$
\begin{eqnarray}
    K_{ij} & = & \sum_{X\in N(A_j)} (2 - d(A_i,X)) + \sum_{X\in N(A_i)} (2 - d(X,A_j)) \label{eq:K_ij}\,. 
\end{eqnarray}
Here, $N(A_i)$ denotes all the main and side chain particles bonded to $A_i$ (e.g., $N(A_i) = \{A_{i-1},A_{i+1},A_i^s\}$ if $i\neq1,N$ and $A_i$ has side chain). 

Note that whenever $A_i$ and $A_j$ are 1-NN, $\lambda_1 (d(A_i,A_j)-1)$ evaluates to $0$. 
Moreover, each neighbor $X$ of $A_j$ either overlaps with $A_i$ and $d(A_i,X) = 0$ or is two turns apart from $A_j$ and $d(A_i,X) = 2$.
Because two turns in the same direction are backtracking, the distance of two lattice sites that are two turns apart can only be $2$ and not $4$.
In the case of no overlaps of the neighboring particles, we find $K_{ij}=0$, yielding $\omega_{ij}\epsilon_{ij}$, which leads to the desired interaction energy $\epsilon_{ij}$ when minimizing. 
If there is an overlap, we have $K_{ij} > 2$, and by choosing $\lambda_2 > -\epsilon_{ij}/2$ for all $i,j$, we can ensure that the contribution is strictly positive and that $\omega_ij$ is set to $0$ when minimizing.

If $A_i$ and $A_j$ are not 1-NN $\lambda_1 (d(A_i,A_j)-1) > 1$ because $d(A_i,A_j) \geq 2$. 
By choosing $\lambda_1$ large enough to compensate for the negative terms from $K_{ij}$, the term inside the summation is strictly positive. 
The bound for $\lambda_1$ has been given in Ref.~\cite{Robert_2021} and requires that $\lambda_1 > 6(j-i + 1) \lambda_2 + \epsilon_{ij}$ for all $i,j$. 

The same approach is applied to interaction pairs of main- and side-chain particles. 
We note that for $A_i$ and $A_j^s$ to be 1-NN, $|i-j|$ has to be even and at least $4$ turns apart, leading to the expression
\begin{equation}
    H_{\rm ms} = \sum_{j\geq i+4\,, (j-i) \text{ even}} ( h^{\rm mcsc}_{ij^s} + h^{\rm mcsc}_{ji^s}   )  \, ,  
\end{equation}
where 
\begin{equation}
    h^{\rm mcsc}_{ij^s} = \omega_{ij^s} ( \epsilon_{ij} + \lambda_1 (d(A_i,A_{j^s}^s)-1) + \lambda_2 K_{ij^s})  \, ,  
\end{equation}
and 
\begin{eqnarray}
    K_{ij^s} =  (2-d(A_i,A_j)) + \sum_{X\in N(A_i)}  (2 - d(X,A_{j^s}^s))  \,. 
\end{eqnarray}
Compared to main-chain particles, where we have at most three 1-NN bonded neighbors ($N(A_i) \leq 3$), side-chain particles have only one 1-NN bonded neighbor, i.e., the corresponding main chain particle, thereby reducing the number of terms in $K_{ij^s}$ compared to $K_{ij}$. 
And while the bound on $\lambda_2$ does not change, the bound on $\lambda_1$ updates to $\lambda_1 > 4(j-i + 2) \lambda_2 + \epsilon_{ij}$ for all $i,j$ because we consider one side-chain particle. 

Finally, the side-side chain interaction term is given by 
\begin{equation}
    H_{\rm ss} = \sum_{j\geq i+3\,, (j-i) \text{ odd}} \omega_{i^sj^s} ( \epsilon_{ij} + \lambda_1 (d(A_{i^s}^s,A_{j^s}^s)-1) + \lambda_2 K_{i^sj^s})  \, ,  
\end{equation}
where 
\begin{eqnarray}
    K_{i^sj^s} =  (2-d(A_{i^s}^s,A_j)) + (2-d(A_i,A_{j^s}^s))  \,. 
\end{eqnarray}
Here, we find that  $\lambda_1 > 2(j-i +3) \lambda_2 + \epsilon_{ij}$. 
The number of required interaction variables is therefore given by
\begin{eqnarray}
    N_{\omega} & = & \frac{(N-5)(N-4)}{4} + \frac{(N-4)(N-3)}{2} + \frac{(N-3)(N-2)}{4}\, 
\end{eqnarray}
and scales $\mathcal{O}(N^2)$, which is why the approach was referred to as ``resource efficient'' in Ref.~\cite{Robert_2021}.

\section{Discussion of $H_{\rm protein}$ and derivation of bounds for penalty parameters}\label{appendix:proofs}

Let us discuss the different terms in $H_{\rm protein}$ and ensure that the Hamiltonian satisfies the necessary requirements. 
Starting with Eq.~(\ref{eq_Hprotein_1}), we consider the relationship between a main chain particle $A_i$ and a protein interaction site $t_l^{I}$.
We consider the situation where $A_i$ is located on the interaction site $t_l^{I}$. 
In this case, $d(A_i,t_l^{I}) = 0$ and we find that the first term reduces to $\epsilon_{il}$.
If all bonded 1-NN of $A_i$ are not on a blocking or invalid site, we have that any pair $(X,Y)$ in the sums in the second term are two turns apart and $d(X,Y) = 2$. 
Hence, the term associated with $\mu_2$ vanishes and we end up, as required, with $\epsilon_{il}$. 
For any bonded 1-NN of $A_i$ on a blocking or invalid site, the distance $d(X,Y)$ evaluates to zero and we add a contribution $2 \mu_2$. 
Hence, by choosing $\mu_2 > -\epsilon_{il}/2$ for all $i,l$, we ensure that the term associated with $\eta_{il}$ in Eq.~(\ref{eq_Hprotein_1}) is strictly positive.

Now let us consider the situation where $A_i$ is not located on the interaction site $t_l^{I}$ and show that the term associated with $\eta_{il}$ is strictly positive. 
We have that the first term is strictly positive whenever $\mu_1 > 1$ because $d(A_i,t_l^{I}) \geq 1$ and $\epsilon_{il} < 0$. 
Next, we note that on the other side the term associated with $\mu_2$ can be negative. 
Hence, we have to  ensure that $\mu_1$ is large enough to compensate any negative contribution of the second term. 
It is sufficient to choose $\mu_1 \geq 1 + 9 \mu_2  |N(A_i)||N(t_l^{I})|/|\epsilon_{il}|$, where $|\cdot|$ applied to sets denotes the cardinality. 
Knowing that $|N_i| \leq 3$ and $|N(t_l^{I})| \leq 3$, we can set $\mu_1 = \mu_2 * 81/|\epsilon_{il}| + 1 $ for all $i,l$.
With proper setting of $\mu_1$ and $\mu_2$, we are confident in the handling of the main-chain particles.

Rather than repeat the above analysis for the side-chain particles $A_i^s$, we just note that the same logic applies, but with a small change. 
That is that the check on the 1-NN neighborhood reduces down to checking the appropriate main chain particle $A_i$ against the blocked and invalid sites, meaning only one summation across $M_{A_i}(t_l^I)$ is needed.
With this, we obtain the expression in Eq.~(\ref{eq_Hprotein_2}).
Finally, the term in Eq.~(\ref{eq_Hprotein_3}) ensures that if for any given interaction site $t_l^I$ more than one associated interaction variable is switched on and set to 1, a penalty term proportional to $\mu_3$ is added. 
Hence, by setting $\mu_3 > 0$ large enough, multiple interaction energies from a single interaction site are penalized and suppressed in the minimization.
In conclusion, we have found that the term $H_{\rm protein}$ with the right choice for $\mu_1$, $\mu_2$ and $\mu_3$ as given above satisfies the required properties. 

It is left to show that the bound 
\begin{equation}
\mu_1 \geq 1+ 9 \mu_2  |N(A_i)||N(t_l^{I})|/|\tilde\epsilon_{il}
\end{equation}
ensures that the terms of $H_{\rm protein}$ in Eq.~(\ref{eq_Hprotein_1}) proportional to the interaction variables $\eta_{ij}$ are strictly positive whenever $A_i$ is not on lattice site $t_l^I$. 
Hence, we require that for any fixed $i,l$ 
\begin{equation}\label{eq:Bound_lambdas}
    \tilde\epsilon_{il}  (1 - \mu_1 d(A_i,t_l^{I})) + \mu_2 \sum_{X\in N(A_i),k} [2 - d(X,t_k^{B})] \geq 0 \, . 
\end{equation}
Solving the equation for $\mu_1$, given that $d_0 := d(A_i,t_l^{I})) >0 $ leads to 
\begin{equation} \label{eq:Bound_lambda1}
   \mu_1 \geq \frac{1}{d_0}  + \frac{ \mu_2} {|\tilde\epsilon_{ij}|d_0} \sum_{X\in N(A_i),k} [d(X,t_k^{B}) - 2] \, . 
\end{equation}
Because $d_0 \geq 1 $, the first term in Eq.~\eqref{eq:Bound_lambda1} can be upper bounded by $1$. 
To upper bound the second term, we note that any $X$ is exactly one turn away from $A_i$ so that its four-dimensional lattice turn representation can be written as $x(A_i)+y$ with $x(A_i)$ given in Eq.~(\ref{eq:turn_representation}), and $y$ a single turn vector with exactly one component equal $1$ and else $0$. 
Similarly, any $t_k^{B}$ can be written as $x(t_l^{I}) + y'$ with $y'$ a single turn vector. Hence, we can bound 
\begin{eqnarray}
     d(X,t_k^{B}) & = &  \sum_i (x(A)_i - x(t_l^{I})_i + y_i - y_i')^2 \\ 
     &  \leq & \sum_i (x(A)_i - x(t_l^{I})_i)^2 + 2\sum_i (x(A)_i - x(t_l^{I})_i)( y_i - y_i') + \sum_i (y_i - y_i')^2 \\ 
      &  \leq &  d_0 + 2 \sqrt{d_0}\sqrt{d(y,y')} + d(y,y') \\ 
      & \leq & d_0 + 4 \sqrt{d_0} + 4
\end{eqnarray}
using the Cauchy-Schwarz inequality in third inequality and $d(y,y') \leq 4$ in the last step. 
Using this inequality, we can upper bound the right hand side of Eq.~(\ref{eq:Bound_lambda1}) by $1 + 9 |N(A_i)||N(t_l^{I})| { \mu_2} /{|\tilde\epsilon_{ij}|}$ using that $d_0 \geq 1$. 
Therefore, choosing $\mu_1$ larger than the derived upper bound suffices to ensure positivity in Eq.~(\ref{eq:Bound_lambdas}) if $ d(A_i,t_l^{I}) > 0$.

\section{QUBO and CP solvers}\label{appendix:solvers}

\subsection{QUBO Reduction}\label{sec:qubo-reduction}

The problem Hamiltonian constructed in Section~\ref{sec:problem} defines an unconstrained binary optimization problem of typically high polynomial order, which we will refer to as a PUBO (for polynomial unconstrained binary optimization).
The higher-order terms are an issue, as current quantum annealing devices (e.g., D-Wave) do not allow for multi-body interactions. Rather, all terms must be quadratic \cite{DWave}.
Because our goal is to construct a solution to the problem that is quantum-amenable---meaning in this case that it could be run on a quantum annealing machine---we must address this restriction.
To this end, we employ a technique known as locality reduction by substitution, wherein an additional variable is introduced and used to replace a two-body interaction term.
However, this comes at a cost: we must also introduce a constraint that this new variable (bit) is equivalent to the product of the pair of variables (bits) it replaced. 
This necessarily adds more variables and terms to the problem Hamiltonian, although those terms are quadratic at maximum.
For each pair of original bits to be replaced $q_i$ and $q_j$, we introduce an ancilla bit $q_{\rm red}$, adding a penalty term $E_{\rm red}$ as
\begin{equation} \label{eq:reduction_hamiltonian}
    E_{\rm red}(q_i,q_j,q_{\rm red},P_{\rm red}) = P_{\rm red}(3q_{\rm red}+q_iq_j-2q_iq_{\rm red}-2q_jq_{\rm red}) , 
\end{equation}
where $P_{\rm red}$ is a penalty weight. 
We can see that $E_{\rm red}=0$ if $q_{\rm red}=q_iq_j$ and $E_{\rm red}=P_{\rm red}$ if $q_{\rm red} \neq q_iq_j$. 
The choice of $P_{\rm red}$ is important, as it must be sufficiently large to prohibit a ground-state solution where $q_{\rm red} \neq q_iq_j$.
This can be tuned carefully according to the problem at hand, or it can be set to cover a worst-case where somehow all ancilla bits $q_{\rm red}$ are misaligned.
In this case, the energy of the Hamiltonian is shifted by as much as the aggregate terms in the expression.
We aligned to this heuristic, and chose our bound as
\begin{equation}
    P_{\rm red}=\sum_{i,j}|Q_{ij}| .
\end{equation}
The optimal choice of substitution bits is an NP-hard problem, and so we leverage existing heuristic-based methods found in the \href{https://pypi.org/project/dimod/}{dimod python library} \cite{Babbush_2014}.
After applying locality reduction, the resulting Hamiltonian is of quadratic order, and is ready to be passed to a solver, such as a simulated or quantum annealing.

\subsection{Hybrid simulated annealing solver}

Simulated Annealing (SA) typically refers to Generalized Simulated Annealing, which combines Classical Simulated Annealing and Fast Simulated Annealing into one optimized framework \cite{Xiang_2013}.
SA is a probabilistic optimization technique useful in solving binary optimization problems, which is what we face here.
In brief, SA operates iteratively over a number of temperature steps $t \in [T_{max}, T_{min}]$.
For any given state (e.g., solution string) $s$, a neighbor state $s'$ is identified, for instance by flipping a bit in $s$.
The energies $E$ of these two states are calculated, and the move to $s'$ is accepted if the move is an improvement [$E(s')<E(s)$], or it is accepted with a temperature-dependent probability $P[E(s), E(s'), t]$ if the move is not an improvement [$E(s')\ge E(s)$].
This probability function $P(\cdot)$ is expected to decrease towards zero as $t \to 0$, at which point the move is only accepted if it is an improvement, turning the algorithm into a greedy descent.
By allowing for moves that do not improve performance, SA provides a mechanism to (potentially) escape local minima, which is a powerful ability in complex optimization energy landscapes.
In this work we leverage the SA solver found in the \href{https://docs.ocean.dwavesys.com/projects/hybrid/en/latest/}{D-Wave hybrid python library}

After SA finds an updated solution to the full problem, the solution is passed to a tabu search solver for additional refinement.
This tabu solver runs what is effectively a greedy neighborhood descent algorithm, taking steps from $s \to s'$ when $E(s') < E(s)$ if those proposed steps $s'$ are not in the tabu list \cite{Beasley_1999}.
The generation of proposed steps $s'$ may vary, but typically are individual or sets of bit flips on variables in $s$.
The tabu list is a fixed sized list of previous visited solutions $s'$, used to prevent immediate solution cycling.
Solutions in the tabu list are removed after some number of steps (i.e., the tenure length) to allow the algorithm to revisit them.
We use the tabu search implementation found in the \href{https://docs.ocean.dwavesys.com/projects/hybrid/en/latest/}{D-Wave hybrid python library}, which is based on Ref.~\cite{Palubeckis_2004}.
After the tabu search has converged on a solution, the loop is restarted and the SA solver starts again.

\subsection{QUBO decomposition}\label{sec:solvers-decomp}

While quantum devices, like the D-Wave Advantage series \cite{DWave}, may house hundreds or thousands of physical qubits, these qubits are not densely connected on the device, and so often multiple physical qubits must be networked together in order to establish long-range connections \cite{Fang_2020}.
This means that individual problem variables (e.g., $q_1$) must be mapped to sets of physical qubits. 
As the problem density (connectivity) increases, more physical qubits must be consumed to enable connectivity, and so fewer problem variables can be mapped onto the device at one time.
If we want to solve large problems, say with hundreds of variables, we will need some mechanism external to the quantum device itself to overcome this limitation.

One way to achieve this is via problem decomposition: Break the full QUBO problem into sub-problems of a given fixed size $k$, solve for the variables in this sub-problem while keeping the other variables fixed, evaluate the newly-generated solution and keep it if it is an improvement, then step to the next sub-problem and repeat until convergence \cite{Rosenberg_2016}.
The choice of sub-problems is key to the success of problem decomposition, and of course there exist numerous generalized options.
For instance, one could choose a starting solution state $\textbf{q}$ and estimate the energy impact of each variable $q_i$ by independently flipping each of their values and measuring the change in system energy.
This creates a mapping of each variable $q_i$ to its measured energy impact on the system, conditioned on the state $\textbf{q}$.
One could then sort the list by descending energy impact and select out $k$ variables at a time to solve for.
This is the default behavior of D-Wave hybrid's \texttt{EnergyImpactDecomposer} object, and is the decomposition strategy we use in this work.
Typically these decomposition strategies assume each variable $q_i$ is included exclusively in a single sub-problem, and once all variables in $\textbf{q}$ have been explored (one full sweep), the process restarts based on the latest best state $\textbf{q'}$; typically $\textbf{q} \neq \textbf{q'}$, else no progress had been made on that sweep.

\subsection{CP Solver} \label{sec:cp-solver}

In this work, we implement the baseline CP model described in Section \ref{sec:problem} using the \href{https://developers.google.com/optimization}{Google OR-Tools CP-SAT solver}. 
The solver employs a variety of primal heuristics and bounding mechanisms and yields provably optimal solutions given sufficient runtime (and valid optimality gaps if the search is terminated early). 
Logical `implies' (e.g., $\rightarrow$) is implemented using the \texttt{OnlyEnforceIf} method, and the logical disjunctions in Eq.~\eqref{core-no-overlap} are implemented as \texttt{BoolOR} constraints. 

\section{Additional CP formulation notes}\label{appendix:cp}
\subsection{RMSD objective function}

In addition to an MJ energy minimizing objective function of the CP model described in Section \ref{sec:problem}, we also explore an alternative objective function that looks to minimize the root-mean-square deviation (RMSD) of our conformation against the true (coarse-grained) coordinates of the peptide, as given by the pipeline described in this paper. 
This RMSD-minimizing objective function yields solutions that more closely match the structure of the real peptide, and can be used to baseline the conformations created with the energy minimizing objective function.

In order to implement this alternative objective function, we introduce a set of new variables $(x^{\delta^2}_{ii_{true}}, y^{\delta^2}_{ii_{true}}, z^{\delta^2}_{ii_{true}})$ where, for example, $x^{\delta^2}_{ii_{true}}$ tracks the squared deviation of the $x$-coordinate of residue $A_i \in A$ with the corresponding true $x$-coordinate of the peptide. 
Then, our RMSD objective seeks to minimize the sum of these across all residues, namely
\begin{equation}
    \min \sum_{i \in A} (x^{\delta^2}_{ii_{true}} + y^{\delta^2}_{ii_{true}} + z^{\delta^2}_{ii_{true}}), 
\end{equation}
as well as for side chain residues. 
The square root is omitted from the objective function due to its negligible impact on the minimization and to reduce the complexity of modeling efforts.

\section{Resource scaling analysis}\label{appendix:scaling}

As part of our initial exploration into solving the peptide docking problem, we have implemented multiple problem formulations in parallel.  
Spatial encoding (see Appendix \ref{appendix:spatial}), the turn ancilla encoding (see Appendix \ref{appendix:turn_ancilla}), and the variable efficient turn encoding (see Sec.~\ref{sec:problem}).
Our goal has been to primarily understand the details of each approach, and also to explore which approach would be able to handle the full peptide and protein active site inclusion. We have implemented feature-complete solutions for each, including all problem constraints and the external protein.
We assess the scalability of these approaches by taking a particular problem instance (in this case, PDB file 2ck0), down-sampling the peptide and protein residues, constructing Hamiltonians for each approach on the down-sampled problems, and measuring the number of variables and terms required for the polynomial unconstrained binary optimization (PUBO) and QUBO Hamiltonians.
We stepped across peptide sizes $N \in [4, 10]$ and $M \in [0, 10]$, with $N=4$ being the minimum length required for a complete cycle (assuming side chains exist) and $N=10$ being the full length of the peptide, and $M$ being the number of protein residues included (with $M=0$ meaning the protein was excluded).
Note that we did not need to include the full set of protein residues to be able to draw conclusions from this analysis.

\begin{figure}[h!]
\centering
\includegraphics[width=0.9\textwidth]{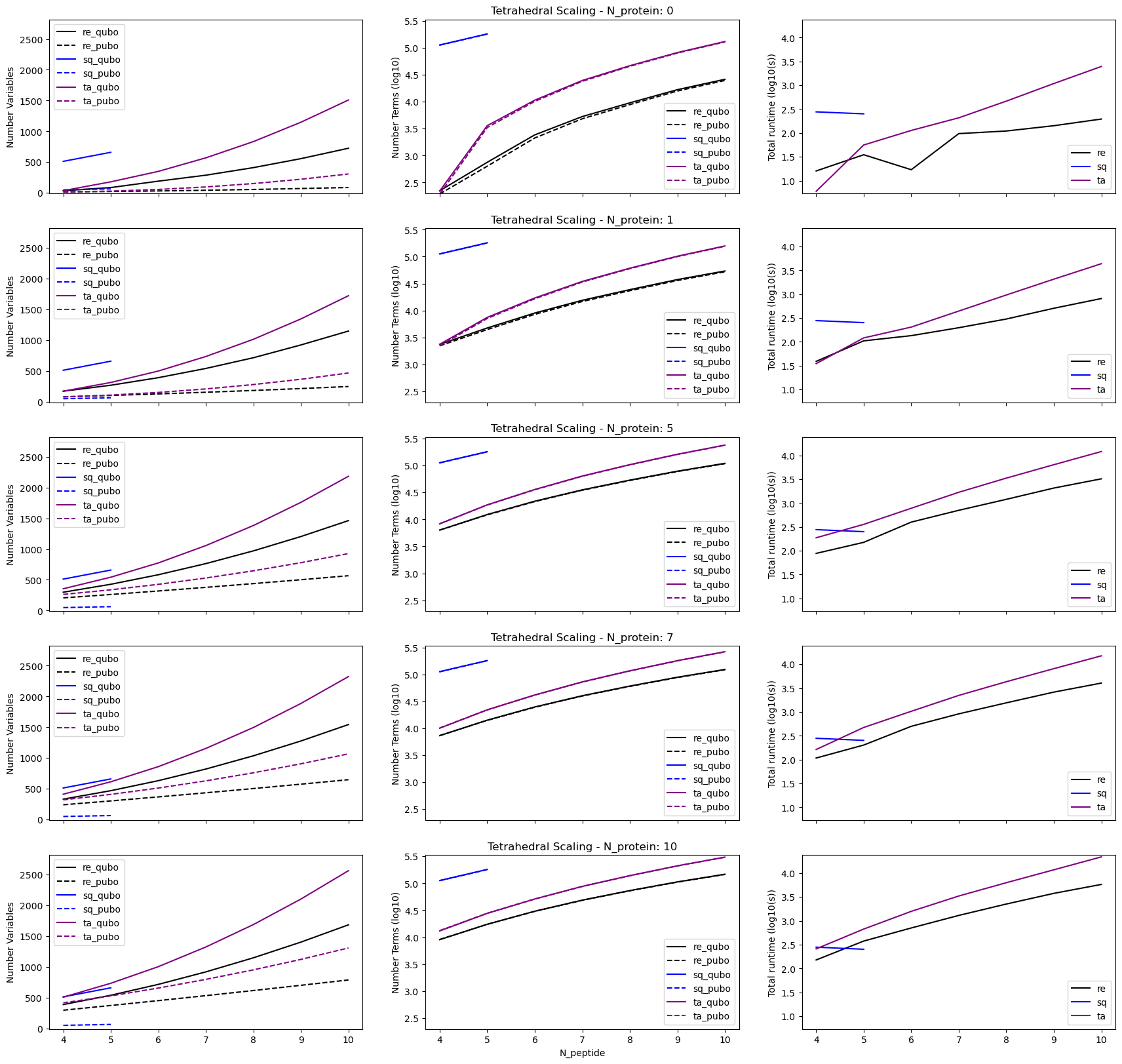}
\caption{Problem scaling for each approach: spatial QUBO (SQ), turn ancilla turn encoding (TA), and resource efficient turn encoding (RE) as a function of the peptide length for polynomial (PUBO) and quadratic (QUBO) Hamiltonian forms. Number of protein residues included starts at $M=0$ (top row) and increases until $M=10$ (bottom row). (left) Number of variables required for the Hamiltonian for each approach. (middle) Number of terms ($\log_{10}$) required for the Hamiltonian for each approach. (right) End-to-end runtime for each approach.}
\label{fig:qubo_pubo_scaling} 
\end{figure}

Figure \ref{fig:qubo_pubo_scaling} shows the results of this analysis. 
We see that the spatial QUBO (SQ) approach is unable to scale beyond $N=5$.
Interestingly, while the number of variables required in the SQ PUBO is the smalles for all approaches, the number of variables in the QUBO is the highest.
This indicates that the SQ Hamiltonian contains terms of very high order, which generate significant overhead in the reduction to a QUBO formulation.
It is expected that reduction from PUBO to QUBO introduces some amount of overhead, in particular in terms of the number of variables in the expression (see Appendix \ref{appendix:solvers}), but it is clear that the SQ approach suffers significantly more from this effect than the resource efficient (RE) or turn ancilla (TA) approaches.
We note that the RE approach---true to its name---is in fact more efficient that either the SQ or TA approach, and this holds across the range of values of $N$ and $M$ tested. 
There is a discontinuity in the number of terms (although less apparent at high $M$) at $N=6$, which is the first length where the main-main cyclization constraint could be enforced, thus adding additional terms to the Hamiltonian.

\section{Spatial formulation}\label{appendix:spatial}

The spatial encoding on the tetrahedral lattice $\mathcal{T}$ is derived from the original two-dimensional (2D) square grid formulation presented in Ref.~\cite{Perdomo_2008}. 
The three-dimensional (3D) tetrahedral lattice is considered a special case of the 3D cubic lattice, where atoms are placed on only opposite corners of each face of a cube, and not all vertices are occupied. 
Here, we adopt and extend the spatial encoding as described in Ref.~\cite{Perdomo_2008}, modifying the Hamiltonian to penalize all the 3D cubic vertices that cannot be occupied so that the resulting lattice is tetrahedral.  

We consider a two-particle (main and side chain) coarse grained (CG) model for each amino acid except for smaller amino acids e.g., glycine, which is represented using one-particle (main chain) coarse grained model. 
The position of each of $N + N^s$ particles (sum of main and side chain) in a D-dimensional cubic lattice can be encoded by \(D(N+N^s)\log_2 (N+N^s)\) binary variables. 
We used the same distance function as described in Eq.~(22) in Ref.~\cite{Perdomo_2008} to represent the rectilinear distance (L1) squared between grid points. 
Note that this distance squared is equal to 1 in square and cubic lattices but is equal to 3 for the tetrahedral lattice. 

Here we describe the logic more symbolically  than in Ref.~\cite{Perdomo_2008}, imagining there is a function $f(*)$ that takes in a peptide particle $i$ (or an equivalent protein-influenced vertex $t^I_l \in \mathcal{T_I}$), and retrieves the bits describing the location of that particle in grid coordinates, in an ordered fashion such that the bit strings of multiple particles $f(i)$ and $f(j)$ can be compared directly. 

\subsection{Hamiltonian construction}

We start our Hamiltonian formulation based on the simplified approach for a 2D grid lattice of one-particle peptide sequence as described in Ref.~\cite{Perdomo_2008}, without cyclization and peptide-protein interaction terms. 
The Hamiltonian is given by
\begin{equation}\label{eq:sq_H_old}
H = H_{\rm onsite} + H_{\rm psc} + H_{\rm pairwise}, 
\end{equation}
where $H_{\rm onsite}$ is an onsite repulsion term for amino acids occupying the same grid point, $H_{\rm psc}$ is a primary sequence constraint term, and $H_{\rm pairwise}$ is a pairwise interaction term that represents favorable hydrophobic interactions between adjacent hydrophobic amino acids. 
Note that original formulation was derived for HP (Hydrophobic-Polar) interactions between amino acids. 
We modify the Hamiltonian and included additional terms, as follows: 
\begin{equation}\label{eq:sq_H0}
H = H_{\rm onsite} + H_{\rm psc} + H_{\rm pairwise} + H_{\rm cycle} + H_{\rm extpairwise} + H_{\rm extonsite} + H_{\rm invalid}, 
\end{equation}
where $H_{\rm cycle}$ is a cyclization constraint term, $H_{\rm extpairwise}$ is a pairwise interaction term between external protein and peptide, and $H_{\rm extonsite}$ is an onsite (steric blocking) constraint term between external protein and peptide.

\subsubsection{Overlap constraint penalty term}

We Eq.~(20) in Ref.~\cite{Perdomo_2008} as the first term in Eq.~(\ref{eq:sq_H0}) to prevent two or more particles (main or side chain) from occupying the same grid point. Symbolically, we can describe the logic as:
\begin{equation}\label{eq:sq_H_onsite}
H_{\rm onsite} = \lambda_0 \sum_{ij} \prod_k {\rm XNOR}]f(i)_k, f(j)_k],
\end{equation}
where ${\rm XNOR}$ is the exclusive NOR operator, returning 1 if both bits are equal, and 0 otherwise. 
In other words, this term steps through all bits in $f(i)$ and $f(j)$ and compares them, and if any bits do not match, the inner product returns 0, otherwise it returns 1, thereby signaling an overlap. 
This is repeated for all pairs of peptide particles $i$, $j$, the overlaps are summed, and a penalty of $\lambda_0$ is applied.

\subsubsection{Primary sequence constraint penalty term}

Extending to a two-particle model mans updating $H_{\rm psc}$ from the original implementation to include the distance requirement between between main and side chains of the same amino acid, i.e., 
\begin{equation}\label{eq:sq_H_psc}
H_{\rm psc} = \lambda_{\rm psc} \left[d^2_{ij}-d'((N + N^s) - 1)\right], 
\end{equation}
where $\lambda_{\rm psc}$ is a strictly positive scalar to regulate the penalty strength, $(N + N^s)$ is the number of main- and side-chain particles, and $d'$ is the distance cutoff of a specific lattice type ($d'=3$ for tetrahedral and $d'=1$ for the square or cubic lattice), and $d^2_{ij}$ represents the squared rectilinear distance between particles $i$ and $j$.

\subsubsection{Pairwise interaction term}

We extend $H_{\rm pairwise}$ to include a MJ potential \cite{Miyazawa_1985} and modify the formulation based on the distance between any two particles on the grid. 
The non-bonded amino acid particles that are one bond distance square apart can have pairwise interactions. 
In 2D  or 3D, one bond distance squared is given by one rectilinear unit distance squared but for the tetrahedral geometry, one bond distance squared is given by 3 rectilinear units distance squared. The term is given by:
\begin{equation}\label{eq:sq_H_pairwise}
H_{\rm pairwise} = \sum_{ij} \Omega_{ij}\left[G_{ij} * (d' - d^2_{ij}) \right] , 
\end{equation}
where $\Omega_{ij}$ is an ancillary bit to turn on or off the potential pairwise interaction between particles $i$ and $j$. 
$G_{ij}$ is the interaction (MJ) energy, $d'$ is the distance cutoff for a specific lattice type ($d'=4$ for tetrahedral and $d'=2$ for square or  cubic lattices), and $d^2_{ij}$ represents the rectilinear distance squared between particles $i$ and $j$.

\subsubsection{Cyclization constraint penalty term}

Next, we add a penalty term $H_{\rm cycle}$ that enforces the cyclization bond $C$ between given peptide residue pairs $i$ and $j$. 
We enforce this constraint by forcing the particles of the residue $i$ and $j$ to be nearest neighbors on the lattice. 
This can be enforced by a penalty term of the form 
\begin{equation}\label{eq:sq_H_cycle}
H_{\rm cycle} = \lambda_{\rm cycle} \sum_{ij} (d^2_{ij} - d'), 
\end{equation}
where $\lambda_{\rm cycle}$ is a strictly positive scalar to regulate the penalty strength and $d'$ is the distance cutoff of a specific lattice type ($d'=3$ for tetrahedral and $d'=1$ for square or cubic lattices), and $d^2_{ij}$ represents the rectilinear distance square between particles $i$ and $j$. 
$H_{\rm cycle}$ is equal to zero if the particles are one lattice step apart and adds a positive contribution otherwise. 
By choosing $\lambda_{\rm cycle}$ appropriately, we can ensure that the minimization leads to a solution satisfying the cyclization constraint.

\subsubsection{Protein interaction terms}

Next, we follow a similar protocol for the protein-peptide interaction as described in Sec.~\ref{sec:problem}. 
However, we first must project the external protein influence onto the lattice vertices and calculate the proper set of influenced coordinates $\mathcal{T_I}$. 
We first introduce the pairwise interaction term between the peptide and external protein:
\begin{equation}\label{eq:sq_H_extpairwise}
H_{\rm extpairwise} = \sum_{i,l} \left[\Omega_{il} G_{il} (d' - d^2_{it^{\mathcal{I}}_l})\right], 
\end{equation}
where $\Omega_{il}$ is an ancillary bit to turn on or off the potential pairwise interaction between peptide particle $i$ and protein interaction-influenced lattice vertex $t^\mathcal{I}_l \in \mathcal{T_I}$. 
$G_{il}$ is the MJ interaction potential, $d'$ is the distance cutoff of specific a lattice type ($d'=4$ for tetrahedral and $d'=2$ for square or cubic lattices), and $d^2_{it^{\mathcal{I}}_l}$ represents the rectilinear distance squared between $i$ and $t^\mathcal{I}_l$.

Next we introduce the onsite term between the peptide and external protein, penalizing overlaps. 
We again do this on the symbolic level, as we did with Eq.~(\ref{eq:sq_H_onsite}), although here we check peptide particles $i$ against protein blocked sites $\mathcal{T_B}$:
\begin{equation}\label{eq:sq_H_extonsite}
H_{\rm extonsite} = \lambda_0 \sum_{il} \prod_k {\rm XNOR}\left[f(i)_k, f(t^I_l)_k\right], 
\end{equation}
again using ${\rm XNOR}$ checks across the bit strings representing the lattice locations of peptide $i$ and protein-blocked site $t^B_l \in \mathcal{T_B}$. 
Note that we use the same overlap penalty term $\lambda_0$.

\subsubsection{Invalid grid points penalty term}

Finally, we need to add an additional Hamiltonian term to penalize the invalid positions on a cubic grid that are not part of the tetrahedral lattice. 
Invalid lattice coordinates are penalized by running overlap checks between positions of all particles and invalid lattice coordinates. 
Symbolically, we can establish the set of valid lattice coordinates, which is equivalent to the tetrahedral lattice set introduced in Section \ref{sec:problem} $\mathcal{T}$, the full set of cubic lattice coordinates as $\mathcal{T_{cubic}}$, and the set of invalid lattice coordinates as
\begin{equation}\label{sq:T_invalid}
\mathcal{T}_{\rm invalid} = \mathcal{T}_{cubic} - \mathcal{T}. 
\end{equation}
With these set definitions, we can check for invalid grid placements by counting the overlaps between each peptide particle $a_i$ and $\mathcal{T}_{\rm invalid}$, as follows 
\begin{equation}\label{sq:H_invalid}
H_{\rm invalid} = \lambda_{\rm invalid} \sum|a_i \cap \mathcal{T}_{\rm invalid}| \ \ \ \  \forall a_i \in A \cup A^s .
\end{equation}
Note that the choice of $\lambda_{\rm invalid}$ is described below. 
In reality, we check the bit strings representing the coordinates in $a_i$ and $\mathcal{T}_{\rm invalid}$, but this is a matter of mapping from the coordinates to the bit strings, so the logic remains consistent.

\subsection{Choice of Penalty terms}

We have seven terms in the final Hamiltonian in Eq.~(\ref{eq:sq_H0}) including pairwise functions. 
$H_{\rm pairwise}$ and $H_{\rm extpairwise}$ are MJ interaction terms, which the model seeks to minimize, as the MJ values are negative.
We assume the same penalty $\lambda_0$ for $H_{\rm onsite}$ and $H_{\rm extonsite}$ as they represent similar overlap constraints, and the same penalty $\lambda_1$ for $H_{\rm psc}$ and $H_{\rm cycle}$ as they represent similar sequence constraints. 
We use $\lambda_{\rm invalid}$ to represent the penalty term for invalid grid points in $H_{\rm invalid}$. 
The choice of penalty terms requires some considerations to achieve optimal solution using classical simulated annealing solvers. 

First, let us consider onsite (penalty) and pairwise (favored) interactions. 
For the square grid, the onsite penalty must be more than three times the maximum of the MJ potential, i.e., $\lambda_0 > 3 \times 7.73 \approx 23$.
For the cubic lattice, the onsite penalty must be more than
five times the maximum of the MJ potential, i.e., $\lambda_0 > 37$. Finally, for the tetrahedral lattice we need to set the onsite and invalid lattice site penalty to be at least five times the maximum of the MJ potential, i.e., $\lambda_0 = \lambda_{\rm invalid} > 37$.
The primary sequence penalty term is given by $\lambda_{\rm psc} = -(n - 1) + d^2_{ij}$ for the square and cubic lattices. For the tetrahedral lattice we use $\lambda_{\rm psc} = -3(n - 1) + d^2_{ij}$.

If all particles were placed on top of each other (i.e.,  the extreme case) the bit strings would either be $[0,0,0,\ldots]$ or $[1,1,1,\ldots]$. The number of onsite violations is therefore $n!/(n-2)!= n(n-1)/2$. Because in this case $d^2_{ij} = 0$ for all $i$ and $j$, the number of primary sequence violations is $(n-1)$. As such, 
$\lambda_0 n(n-1)/2$ must be larger than $\lambda_1(n-1)$.
In this work we treat the primary sequence and cycle penalties in the same way, i.e., if all particles would be restricted to one lattice vertex there will be favorable energy from $H_{\rm cycle}$. 
For the square and cubic lattices this energy is $(d^2_{ij} - 1)$ and for the tetrahedral lattice $(d^2_{ij} - 3)$. 
Therefore, $\lambda_0 n(n-1)/2$ must be larger than $\lambda_1 (n-1+1)$.
This is the extreme case for the 3D lattice. For the tetrahedral lattice, $\lambda_0 n(n-1)/2$ must be more larger than $3\lambda_1(n-1+1)$. Setting $n=2$, we obtain $\lambda_0 > 6\lambda_1$. 
It is probably safe to set $\lambda_0 = 10\lambda_1$.
However, because $\lambda_0 > 37$, $\lambda_1 \ge 4$.
Finally, if we consider the onsite penalty and invalid site penalty for the tetrahedral lattice, one option is to keep $\lambda_0 = \lambda_{\rm invalid}$. However, the avoid violations we set  $\lambda_{\rm invalid} = 2\lambda_0$. Summarizing, we use the following values for the different penalty terms: $\lambda_1$ = 4, $\lambda_0$ = 10 $\lambda_1 = 40$, and $\lambda_{\rm invalid}$ = 2 $\lambda_0 = 80$.

\section{Turn ancilla formulation}\label{appendix:turn_ancilla}

Here we describe an alternative turn encoding approach---the turn ancilla encoding---which is an extension of the work of Babbush \textit{et al.}~in Ref.~\cite{Babbush_2014}.
We seek to formulate the problem described as a PUBO, which can then be reduced to a QUBO and solved using an optimizer such as simulated annealing.

The turn encoding defines the peptide conformation by a series of turns, either along the main-chain backbone, or off of it to place side-chain particles, starting from the first main-chain residue (assumed to be placed at the lattice origin).
The tetrahedral lattice has $4$ possible turns from any given vertex, which are counter-directional on alternating sublattices: $t \in (0,1,2,3)$ on $\mathcal{T_A}$ and $t \in (\bar{0},\bar{1},\bar{2},\bar{3})$ on $\mathcal{T_B}$.
We leverage a bit mapping that maps  $(0,1,2,3) \mapsto (00,01,10,11)$ on $\mathcal{T_A}$, and $(\bar{0},\bar{1},\bar{2},\bar{3}) \mapsto (00,01,10,11)$ on $\mathcal{T_B}$.
For a peptide sequence $A$, we require $|A|-1$ turns for the main chain and $|A^s|$ turns for the side chain, which, requiring 2 bits per turn, means we need a total of $|q| = 2*(|A|-1+|A^s|)$ bits to define the peptide conformation.
Distance between particles on the lattice is defined by  Eq.~(\ref{eq:tetra_dist}). 
As a reminder, the distance between a particle and a lattice vertex can be calculated by the same equation, noting that in the case of a vertex, the turn vector $x$ would represent the number of steps along each dimension $k$ from the origin to that vector.
With this in mind, we can now discuss the turn ancilla Hamiltonian construction.

\subsection{Turn ancilla Hamiltonian}

We start by reviewing the Hamiltonian terms introduced in Ref.~\cite{Babbush_2014}, although we note that Babbush {\em et al.}~built the peptides on a square grid whereas we are building these on a tetrahedral lattice. Thus,  we  update the definitions of these Hamiltonian terms accordingly
\begin{equation} \label{eq:ta_H0_old}
H = H_{\rm back} + H_{\rm pair} + H_{\rm overlap}, 
\end{equation}
where $H_{\rm back}$ is a special case of $H_{\rm overlap}$, checking that sequential turns do not yield overlapping residues (i.e.,   backtracking), $H_{\rm pair}$ calculates the total MJ interaction potential between relevant residue pairs in the solution, and $H_{\rm overlap}$ calculates a penalty for residue pairs that overlap on the same lattice vertex.
To define $H_{\rm back}$, we use the helper function $T(t,t') = \sum_{k=0}^3 f_k(t)f_k(t')$, which, again, is equal to $1$ if turns $t$ and $t'$ are in the same direction and $0$ otherwise.
We can then build $H_{\rm back}$ using the helper function as follows:
\begin{equation} \label{eq:ta_H_back}
H_{\rm back} = \sum_{i=1}^{N-1} T(t_i,t_{i+1}) + \sum_{i=2}^{N-1} T(t_{i-1},s_{i}) + \sum_{i=1}^{N-1} T(t_{i},s_{i}) \, ,
\end{equation}
which checks for backtracking on sequential main-chain turns, on the previous main-chain turn and current side-chain turn, and on the current main-chain turn and side-chain turn.

To understand $H_{\rm pair}$, we note that $g_{ij}$ in Eq.~\eqref{eq:ta_H_pair} is used to represent the squared distance between residues $i$ and $j$; in other words, it represents the value of Eq.~(\ref{eq:tetra_dist}).
As part of our formulation, we assume that only residue pairs at a distance of 1 (i.e., 1-NN) can interact with one another. Thus, we need an equation for each residue pair that returns a negative value (i.e.,  is beneficial) when $g_{ij} = 1$ and $0$ otherwise. 
Noting that $g_{ij} \geq 0$, we can accomplish this via
\begin{equation} \label{eq:ta_H_pair}
H_{\rm pair}(i,j) = \omega_{ij}  {\rm MJ}(i,j) (2-g_{ij}) , 
\end{equation}
where $\omega_{ij}$ is the interaction ancilla for pair $(i,j)$ and ${\rm MJ}(i,j)$ is the Miyazawa-Jernigan potential between pair $(i,j)$.
If $g_{ij} = 0$, then we have an overlap between particles, which should be penalized via $H_{\rm overlap}$, and if $g_{ij} > 2$ then $2-g_{ij}$ is negative, and because ${\rm MJ}(i,j)$ is always negative, the result is a positive value, which the solver should avoid. 
The interaction ancillas $\omega_{ij}$ are useful in preventing the solver from adding pairwise interaction terms between all main chain residues (i.e., residues that are connected by a turn), as those residues will always be a distance of 1 from each other. 
As such, $\omega_{ij}$ prevents the solver from interpreting these residue pairs as beneficial to the overall solution.

$H_{\rm overlap}$ penalizes any residue pairs that are at a distance of 0 (i.e., overlapping), but to do so in a manner that leverages a fixed number of variables, which can be specified prior to solving. 
To do this, rather than compare the distance between two residues $g_{ij}$ against 0 (which is an unbounded comparison), we compare the distance against the upper bound of that distance $2^{mu_{ij}}$. 
This is why we require slack ancillas: We want to ensure the slack can compensate for distance up to $2^{mu_{ij}}-1$, so that if $g_{ij} \geq 1$, then $H_{\rm overlap} = 0$, else if $g_{ij} = 0$, then $H_{\rm overlap} = 1$.  We start top-down, specifying that for each potentially interacting residue pair $i,j$
\begin{equation} \label{eq:ta_gamma_ij}
\gamma_{ij} = (2^{\mu_{ij}} - g_{ij} - \alpha_{ij})^2 , 
\end{equation}
where $2^{\mu_{ij}}$ is the maximum possible squared distance between residues $(i,j)$ (i.e., if they were in a straight line) and $\alpha_{ij}$ specifies the ancilla bits relating to residue pair $(i,j)$. Note that 
\begin{equation} \label{eq:ta_alpha_ij}
\alpha_{ij} = \sum_{k=0}^{\mu_{ij}-1}q_{c_{ij}+k}2^{\mu_{ij}-(k+1)} \,\,\,\leq 2^{\mu_{ij}} - 1 , 
\end{equation}
where $c_{ij}$ is a pointer to the correct starting bit for $\alpha_{ij}$, which requires $k$ bits, among all bits in the solution string.
Finally, when applied to the potentially interacting residue pairs, we obtain
\begin{equation} \label{eq:ta_H_overlap}
H_{\rm overlap} = \sum \gamma_{ij}
\end{equation}
which returns the total number of overlaps, or 0 if there are none. 
This result is then scaled by a penalty weight $\lambda_{\rm penalty}$.

\subsubsection{Cyclization constraint term}

We define the cyclization constraint as a constraint enforcing the distance between the residue pairs specified in $C_{ij}$ to be equal to $1$. 
In other words, $g_{ij} = 1 \forall i,j \in C$. 
Because we have a formula for the distance between any two residues, this constraint is easily implemented. 
We add a term to the system Hamiltonian $H_{\rm cycle}$, which encodes this constraint:
\begin{equation} \label{eq:ta_H_cycle}
H_{\rm cycle} = \lambda_{\rm penalty} \sum_{C}(g_{ij} - 1)^2 , 
\end{equation}
where $\lambda_{penalty}$ is a penalty weight.
The choice of $\lambda_{\rm penalty}$ is important, because using a value that is too small may (and likely will) yield solutions that violate these constraints. However, values too large may prevent the solver from exploring infeasible regions on its way to finding potentially optimal regions. To find a minimum useful value for $\lambda_{\rm penalty}$, we can calculate the energy of a solution where all peptide residues occupy two lattice vertices, which would yield the maximal intra-peptide energy $E_{\rm peptide}$. Alternatively, the peptide could form a straight, non-cyclized chain, such that is maximizes the peptide-protein interaction energy $E_{\rm peptide-protein}$. 
These values vary depending on the problem configuration, but once calculated, we can consider setting $\lambda_{\rm penalty} = \max\{|E_{\rm peptide}|, |E_{\rm peptide-protein}|\}+1$. In practice, we find that setting the value much higher, for instance to $lambda_{\rm penalty}=1000$ to be simpler and more useful to keep the model results from violating these constraints.

\subsubsection{Protein interaction term}

To derive the protein interaction  term, we first project the protein particles' effect onto the lattice, yielding a set of blocked vertices $\mathcal{T_B}$ and a set of interaction vertices $mathcal{T_I}$.
In order for a peptide residue $i$ to feel the effect of the protein, it must directly overlap with a projected protein vertex $t_l$ ($d(i,l)=0$).
We expect that any interaction site $t_l^I$ will have some number of 1-NN blocked sites $t_l^B$, and so if residue $i$ does overlap with interaction site $t_l^I$, we must check for 1-NN blocking sites $t_k^B$. The Hamiltonian therefore is given by
\begin{eqnarray} \label{eq:ta_H_prot}
    H_{\rm protein} & = & \sum_{i,l} \eta_{il}  \big( \tilde\epsilon_{il}  (1 - \lambda_1 d(A_i,t_l^{I})) + \lambda_2 \sum_{X\in N(A_i),k} (2 - d(X,t_k^{B})) \big) \\
                &   & + \sum_{i,l} \eta_{i^sl} \big( \tilde\epsilon_{il}  (1 - \lambda_1 d(A_i^s,t_l^{I})) + \lambda_2 \sum_{k} (2 - d(A_i,t_k^{B})) \big) \, .
\end{eqnarray}
For any peptide residue $A_i$ and protein interaction site $t_l^I$, if $d(A_i, t_l^I) > 0$ then there is a positive energy contribution which should be turned off by $eta_{i,l}$, else if $d(A_i, t_l^I) = 0$, then we perform 1-NN checks on $N(A_i)$ and $N(t_l^I)$ to confirm no peptide particles land on a blocking site.
We chose to include the external protein interactions in this manner as opposed to continuing with the turn ancilla bits, because the variable scaling limitation of the turn ancillas before attempting to include the protein.
As we can see in Appendix \ref{appendix:scaling}, the variable scaling of the resource efficient approach for peptide-peptide interactions is of the order $\mathcal{O}(N^2)$, whereas the variable scaling for turn ancilla is of the order of $\mathcal{O}(N^2 \log_2(N))$.

\subsection{Complete Hamiltonian}

Summing over all terms derived above, results in the following composite turn ancillary problem Hamiltonian:
\begin{equation} \label{eq:ta_H0}
    H = H_{\rm pair} + H_{\rm protein} + \lambda_{\rm penalty}(H_{\rm overlap} + H_{\rm back} + H_{\rm cycle}).
\end{equation}

\section{Additional experimental results}\label{appendix:experiments}
\subsection{Experimental Setup}

We have implemented the problem instance generation pipeline and the QUBO and CP solution approach in Python. 
For the QUBO approach, we use the \href{https://github.com/dwavesystems/dimod}{dimod library} from D-Wave Systems to perform the PUBO to QUBO locality reduction. 
We use the open source \href{https://github.com/dwavesystems/dwave-hybrid}{D-Wave Hybrid} framework for the simulated annealing (SA) and decomposition-based experiments. 
For each problem instance (peptide), we run hyperparameter optimization (HPO) via Amazon SageMaker over the following parameters: $\lambda_{\rm cycle}$, $\lambda_{\rm back}$, $\mu_{2}$, $\mu_{3}$, $p$, $sub\_qubo\_size$, $num\_reads$, $num\_sweeps$.  
Because SA is a probabilistic algorithm, we run $10$ shots for each of the problem instances on each fixed parameter solver run. 

For the HPO algorithm, we use the SageMaker Tuning Job built-in Bayesian optimization, the details of which can be found in the \href{https://docs.aws.amazon.com/sagemaker/latest/dg/automatic-model-tuning-how-it-works.html}{SageMaker Documentation}.
Each step in the HPO process reports back only the best loss result across all $10$ shots, and the optimization acts on that value.
Each HPO step is allowed to run for up to $4$ hours, and if that time is exceeded, the step is considered a failure and results are not reported.
We limit the HPO search to $100$ steps, truncating the run if the algorithm has not converged by then.

The CP model is implemented using the OR-Tools library \cite{Perron_2011} and uses all settings at their default values, except for the run-time limit, which is set to 300 seconds. 
The compute resources used for the experiments are Amazon Web Services Inc. (AWS) Elastic Cloud Compute (EC2) \href{https://aws.amazon.com/ec2/instance-types/m5/}{m5.4xlarge} instances. 
These instances feature an Intel Xeon Platinum 8000 series processor, with 16 virtual CPUs and 64GB of memory. 

The visualizations presented in this paper use a combination of the \href{https://www.rcsb.org/3d-view/}{Mol* 3D viewer} available from the RCSB Protein Data Bank as well as \href{http://3dmol.org/}{3dmol.js}. 

\subsection{Pipeline definition}

\begin{figure*}[ht!]
   \subfloat[Split PDB into protein and peptide \label{fig:workflow_frame1}]{%
      \includegraphics[ width=0.5\textwidth]{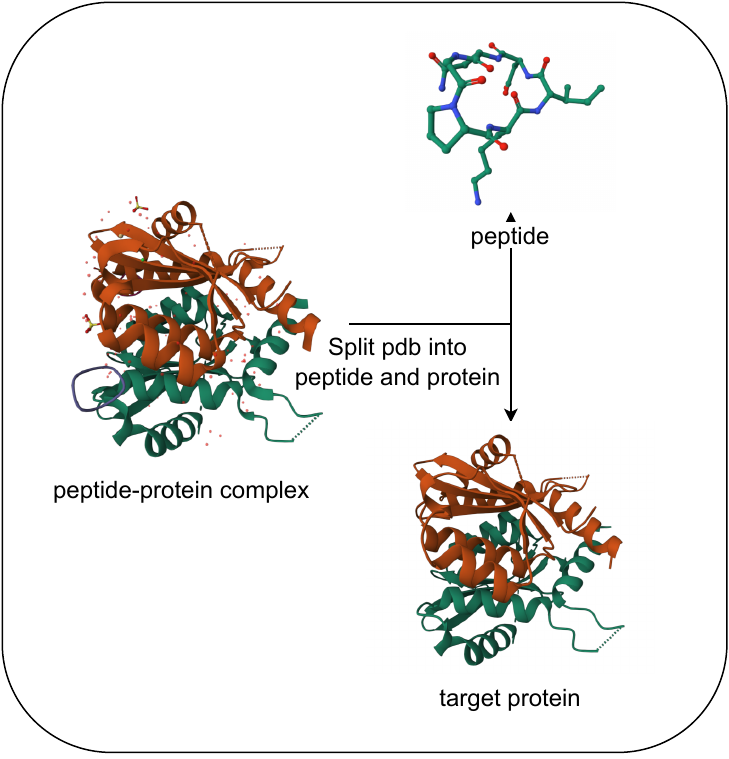}}
   \subfloat[Generate active sites
   \label{fig:workflow_frame2}]{%
      \includegraphics[ width=0.5\textwidth]{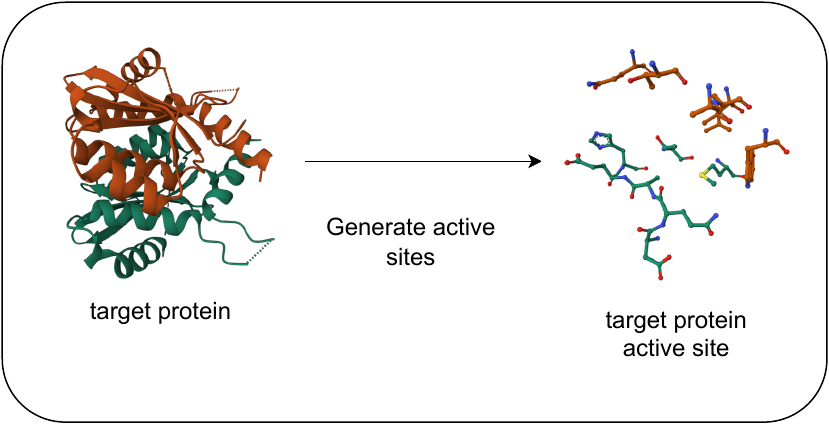}}
\hspace{\fill}
   \subfloat[Convert to coarse grained particles \label{fig:workflow_frame3}]{%
      \includegraphics[ width=0.5\textwidth]{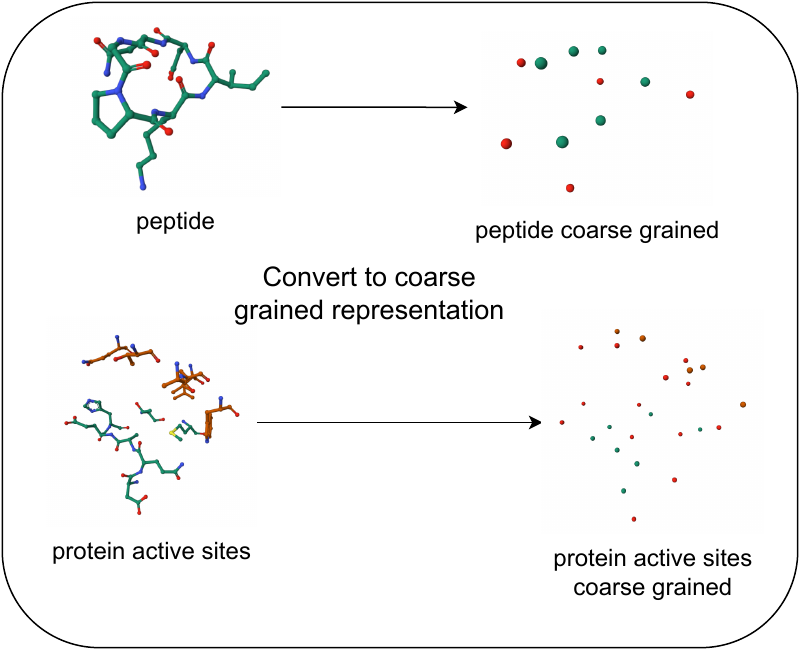}}
   \subfloat[Construct tetrahedral lattice \label{fig:workflow_frame4}]{%
      \includegraphics[ width=0.5\textwidth]{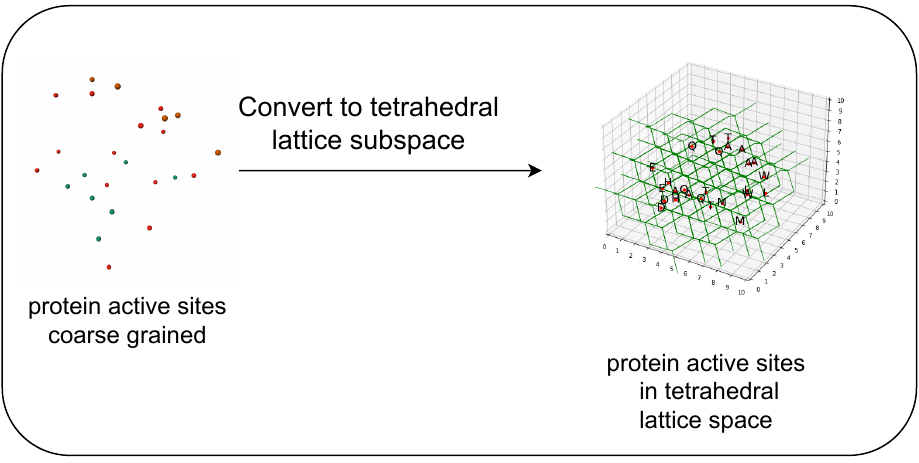}}
\caption{Problem instance generation pipeline for peptide 3WNE from an atomic-level, continuous coordinate PDB representation to a 2-particle coarse grained tetrahedral lattice representation. Details are described in the main text.}
    \label{fig:workflow}
\end{figure*}

Here we detail the pipeline defined in Section \ref{sec:experiments} and outlined in Fig.~\ref{fig:workflow}. For all experiments we use a lattice distance of $3.8\AA$, and for the target protein interaction a blocking radius of $2.5\AA$ to $3.5 \AA$ (steric hindrance), and MJ interaction potentials within a radius of $6.5\AA$.
We note that our solution allows for different radii to be specified, so long as the interaction radius is at least 1.633 times larger than the blocking radius.

The first step of our approach is to separate the information in the PDB file into two components, namely the peptide, and the target protein (i.e., the protein that the peptide docks onto). 
Extracting these components is easily achieved using the information in the header of the PDB file. 
For protein 3WNE specifically, the header specifies that the PBD contains two types of molecules: A polyprotein (i.e., the target protein) and a synthetic peptide (i.e., the peptide we seek to dock on the target). 
Note that these two components are characterized by a series of coordinates specifying where each atom of the protein/peptide is located in space. 
The next step in the procedure is to identify the protein active site (Figure \ref{fig:workflow_frame2}), which represents a smaller, more focused area that the peptide is likely to use for docking. 
Determining an effective active site is important because smaller active sites result in fewer variables in the problem encoding, and thus a higher likelihood of being able to produce a solution. 
To isolate the active site we find the residues in the target protein that are less than some distance threshold (in this case, $5\AA$) from the peptide residues. 
Note that the specific distance threshold used impacts the size of the active site, and thus the quality of the solutions produced by our optimization approach.

With the target protein active site and peptide coordinates in hand, the next step is to calculate the coarse-grained 2-particle representation of each of the amino acids in these structures (Figure \ref{fig:workflow_frame3}). 
This is accomplished by classifying each of the atoms in each residue as main chain or side chain and then calculating the centroid (adjusted by molecular weight) of these groupings. 
The result is a two-particle representation of the amino acid, where each of the particles is characterized by a coordinate and an amino acid label, as described in Section~\ref{sec:problem}. 
Note that in the case of Glycine, only a single particle is generated because there is no side chain. 

The last step in our problem instance generation procedure is to produce the tetrahedral lattice that discretizes the search space for the binary optimization approach (Figure \ref{fig:workflow_frame4}).
Our approach for placing the tetrahedral lattice is described in Algorithm \ref{alg:lattice-gen}. 
We generate an initial candidate for the lattice origin by taking the centroid of the active site $P$ and then grow the tetrahedral lattice recursively up to depth $\delta$.
In our implementation we enforce discrete-valued lattice vertices by rounding the origin coordinates to the nearest integer, and then using an edge length of $\sqrt{3}$ for the tetrahedral lattice. 
We assume that lattice edges correspond to a distance of $3.8\AA$, and so we scale the coordinates of the active site $P$ accordingly (e.g., a protein particle with coordinate $(1\AA,1\AA,1\AA)$ in the original PDB file becomes $({\sqrt{3}}/{3.8}, {\sqrt{3}}/{3.8}, {\sqrt{3}}/{3.8})$ in our tetrahedral lattice scale). 
We estimate the required depth $\delta$ based on the distance from the selected origin to the furthest point in the active site. 
The lattice is then filtered according to the current active site coordinates $P$ and a steric clash parameter $\alpha$ to remove vertices that would clash with the target site, $\mathcal{T_B}$. 
A subroutine tests whether the resultant lattice is capable of producing a feasible conformation (e.g., the origin is checked for steric clash and evidence there is at least one route available to yield a cyclic conformation). If the lattice is invalid, the target site coordinates are shifted away from the origin of the lattice (by step size $\delta$ in the direction of the rest of the target protein) and the process repeats until a valid filtered lattice is determined. 

\begin{algorithm}[tb]
\caption{Construct filtered tetrahedral lattice, $\mathcal{T}' = \{t_1, t_2, \dots \}$}
\label{alg:lattice-gen}
\textbf{Input}: Target site centroid representation, $P$, lattice depth, $\delta$, steric clash radius, $\alpha$, shift step $\delta$ \\
\textbf{Output}: Filtered tetrahedral lattice, $\mathcal{T}'$, shifted target site coordinates, $P$
\begin{algorithmic} 
\STATE $t_1 \leftarrow \textsc{centroid}(P)$ \COMMENT{determine the origin of lattice}\\
\STATE $\mathcal{T} \leftarrow \textsc{growLattice}(t_1, \delta)$ \COMMENT{grow the lattice from the origin according to depth} \\ 
\STATE $\mathcal{T}' \leftarrow \mathcal{T} \setminus \mathcal{T_B}$ \COMMENT{remove vertices affected by steric clash} \\ 
\WHILE{\textsc{invalidLattice}($\mathcal{T}')$}
    \STATE $P \leftarrow \textsc{shift}(P, \delta)$ \COMMENT{shift target site away from lattice origin} \\
    \STATE $\mathcal{T}' \leftarrow \mathcal{T} \setminus \mathcal{T_B}$ \COMMENT{remove vertices affected by steric clash}\\ 
\ENDWHILE
\end{algorithmic}
\end{algorithm}

\subsection{Pipeline run times}

\begin{figure}[h]
\centering
\includegraphics[width=0.9\textwidth]{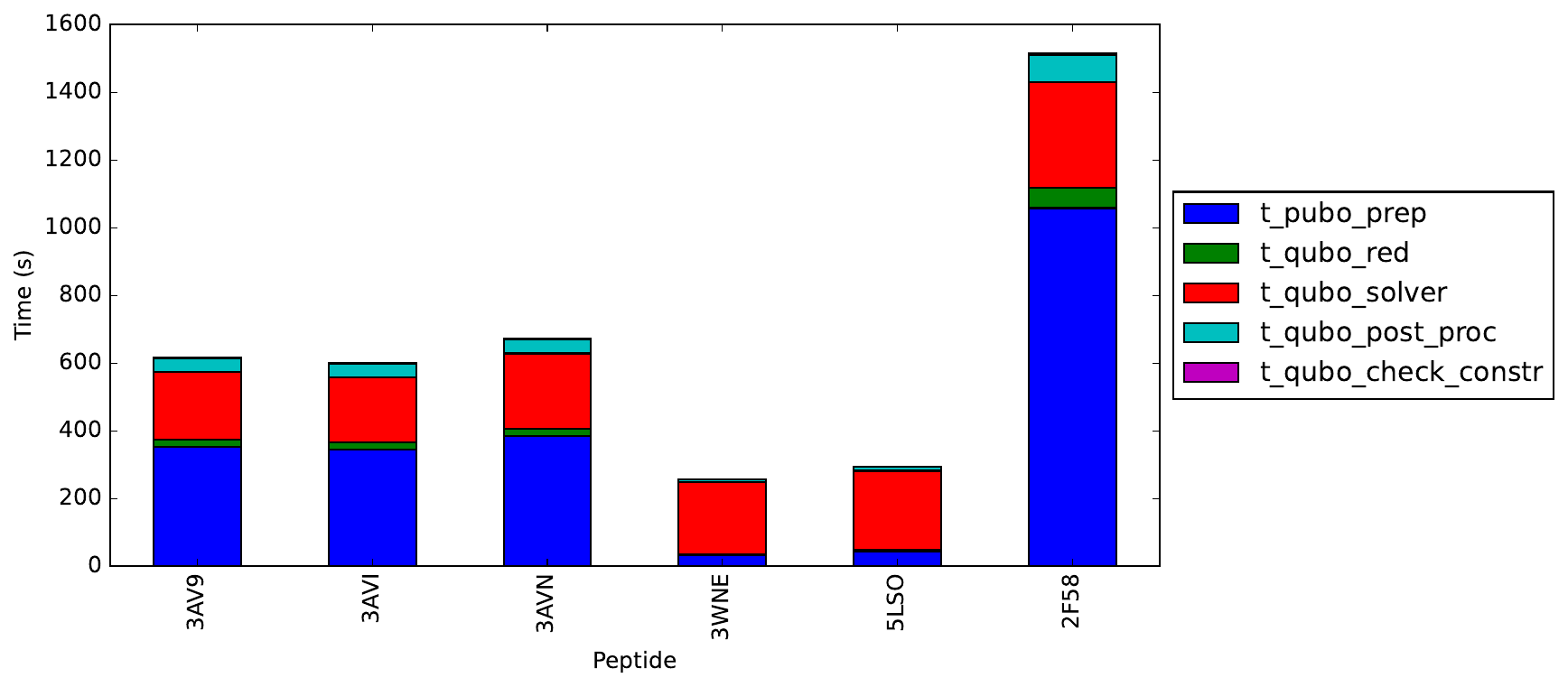}
\caption{Pipeline run time analysis. For each instance, the arithmetic mean of each stage in the pipeline is evaluated over 10 shots. }
\label{fig:pipeline-runtime}
\end{figure}

We visualize the run time of the pipeline applied to the various peptide instances in Figure \ref{fig:pipeline-runtime}. 
Each segment in the stacked bar chart represents the arithmetic mean across the $10$ shots per instance. 
Here, $\texttt{t\_pubo\_prep}$ refers to the time required to build and expand the Hamiltonian while $\texttt{t\_qubo\_red}$ identifies the time required to translate the Hamiltonian into a solver-amenable QUBO formulation. 
$\texttt{t\_qubo\_solver}$ is the actual time spent doing the optimization. 
Finally, $\texttt{t\_qubo\_post\_proc}$ and $\texttt{t\_qubo\_check\_constr}$ refer to routines run after the optimization is complete for post processing and constraint violation checks.
 
Note that the construction of the PUBO Hamiltonian an time-expensive stage of the pipeline, becoming more of a burden as peptide size increases. 
The run time in this stage is largely dominated by the need to mathematically expand the large number of terms of the produced Hamiltonian before the quadratic transformation. 
The number of terms generated scales quadratically with the number of peptide particles, as each must be compared with all other peptide particles, and with all protein particles. 
In our implementation, we use the \texttt{sympy} library to accomplish these steps. 
The PUBO to QUBO reduction process takes considerably less time 
and the QUBO solver run time is comparable across instances, ranging from approximately 3.5 min for 5LSO to approximately 6.5min for 3AVI.

\subsection{MJ energy and RMSD}

\begin{figure}[ht!]
\centering
\includegraphics[width=0.9\textwidth]{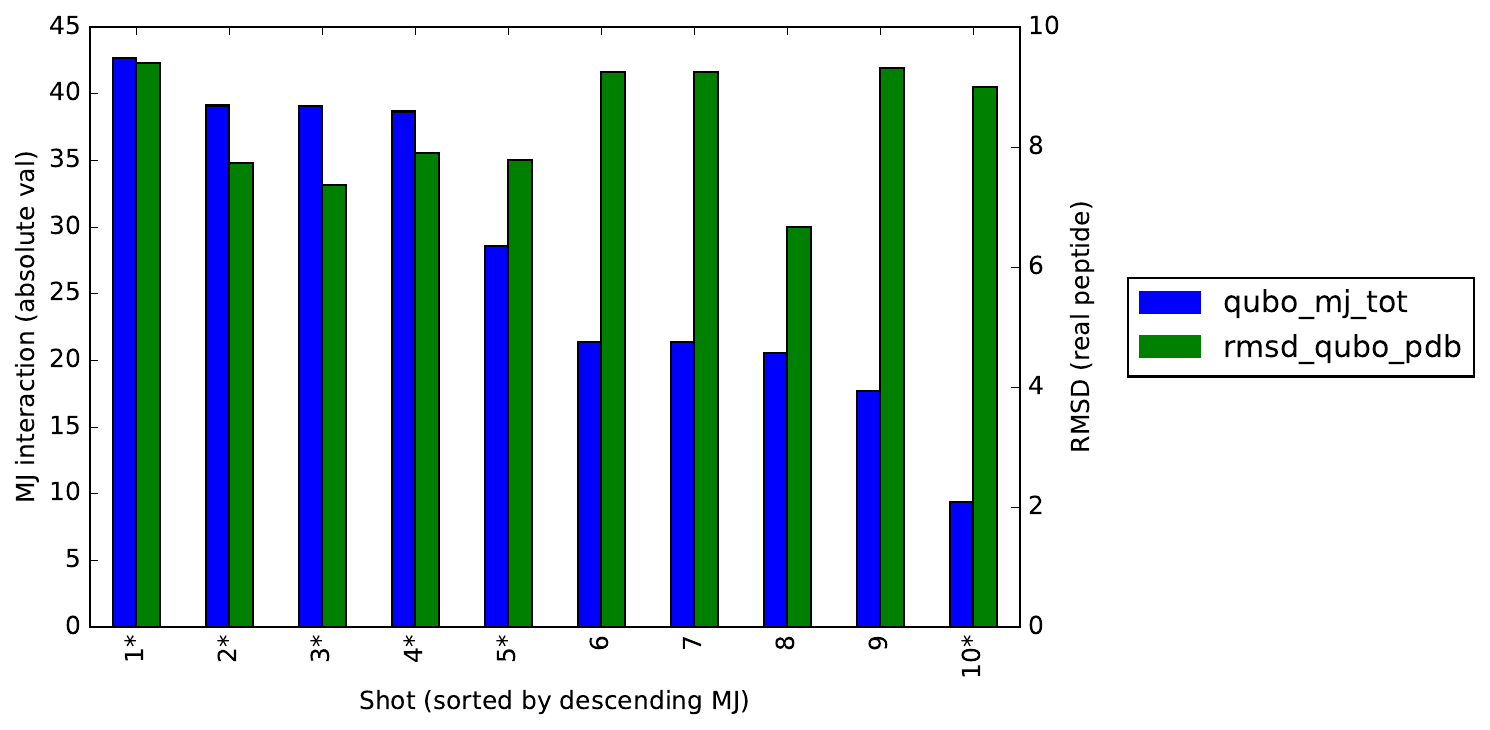}
\caption{Shots for the 3WNE peptide sorted by descending absolute MJ value in blue (the higher the better). Corresponding RMSD with real peptide shown in green. Infeasible solutions marked with asterisk (*).}
\label{fig:qubo_mj_rmsd} 
\end{figure}

The core underlying assumption of our modeling approach is that by optimizing the aggregate MJ potentials, we necessarily find the highest quality solution, which should equate to the true PDB result. 
Our results suggest this might not be the case, or perhaps that MJ 
potentials alone are insufficient.
Figure \ref{fig:qubo_mj_rmsd} illustrates the potential weakness in using an MJ objective as a proxy for RMSD. 
Here we sort the 10 shots for the 3WNE peptide by descending MJ energy (i.e., the left-most bar corresponds to the best MJ solution shot), and show the RMSD value for the solution against the real peptide. 
It is apparent that the solution with the best MJ energy does not correspond to the solution with the lowest RMSD to the real peptide; in fact, it would appear that the lowest RMSD solution has the second-best MJ value among the feasible solutions. 
This is because the solver is incentivized to find solutions that have high MJ interactions, but not necessarily to place residues in close proximity to the real peptide because at the time of solve, the real peptide coordinates are not known.
It is important to note that these results are derived from the QUBO model, which is anchored at the lattice origin. 
While we expected this to introduce some error, we also expected to see that a result with a higher aggregate MJ potential has also lower RMSD.
Instead, we do not see a strong relationship between aggregate MJ and RMSD.

\subsection{Additional Solution Visualizations}

We provide additional three-dimensional visualizations of CP and QUBO solutions for the remaining peptides in Table \ref{table:problem-instances}, to help give context to the results.

\begin{figure}[ht!]
\centering
\includegraphics[width=0.9\textwidth]{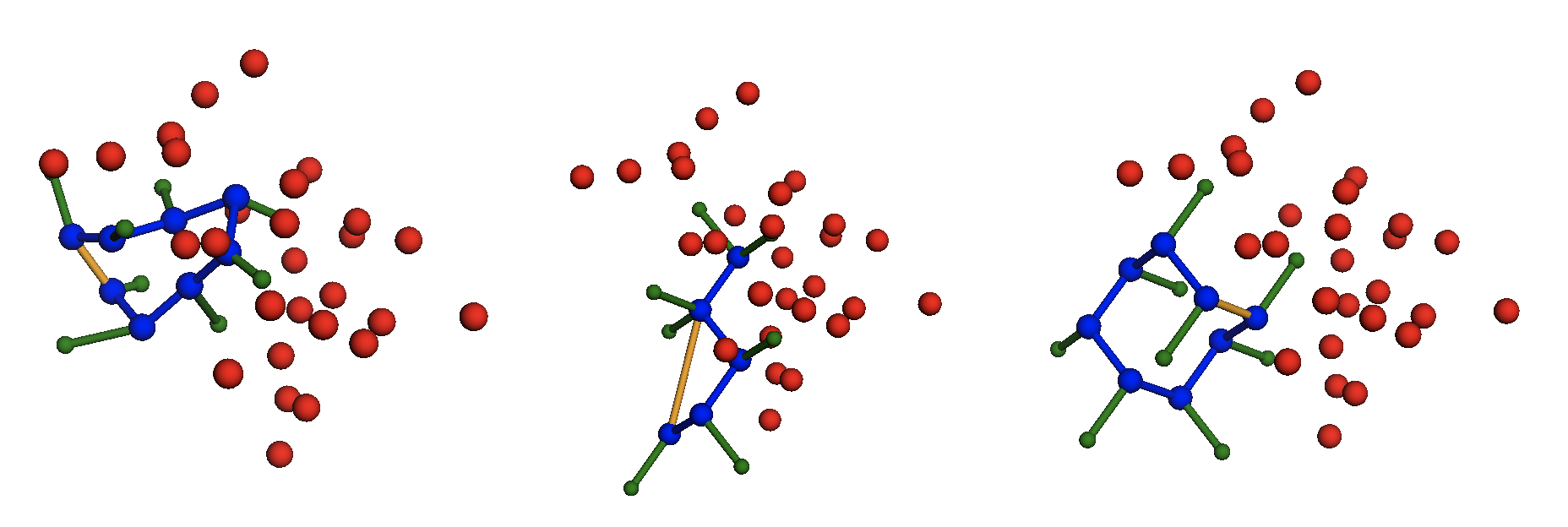}
\caption{3AVN result visualization. True PDB peptide (left) against the QUBO model result (middle) and the CP-MJ 1-NN result (right). The peptide is represented with blue and green dots (main and side-chain, respectively, conected by lines), and the red dots are the external protein residues.}
\label{fig:3avn-result-comparison-plot} 
\end{figure}

\begin{figure}[ht!]
\centering
\includegraphics[width=0.9\textwidth]{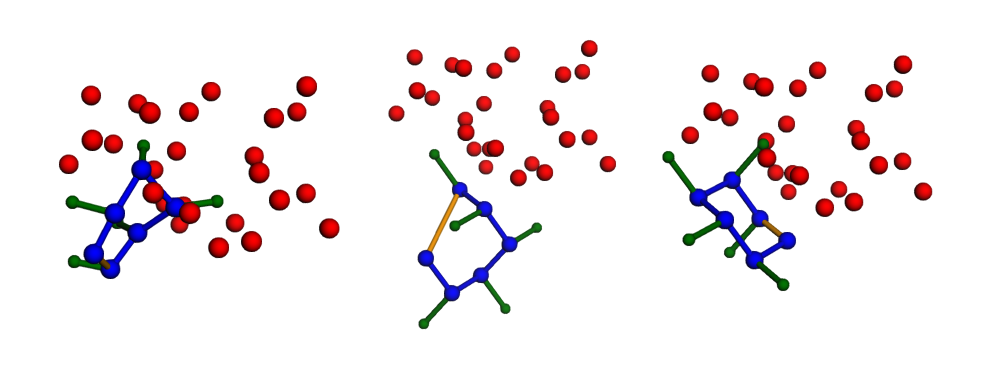}
\caption{3WNE result visualization. True PDB peptide (left) against the QUBO model result (middle) and the CP-MJ 1-NN result (right). The peptide is represented with blue and green dots (main and side-chain, respectively, connected by lines), and the red dots are the external protein residues.}
\label{fig:3wne-result-comparison-plot} 
\end{figure}

\begin{figure}[ht!]
\centering
\includegraphics[width=0.9\textwidth]{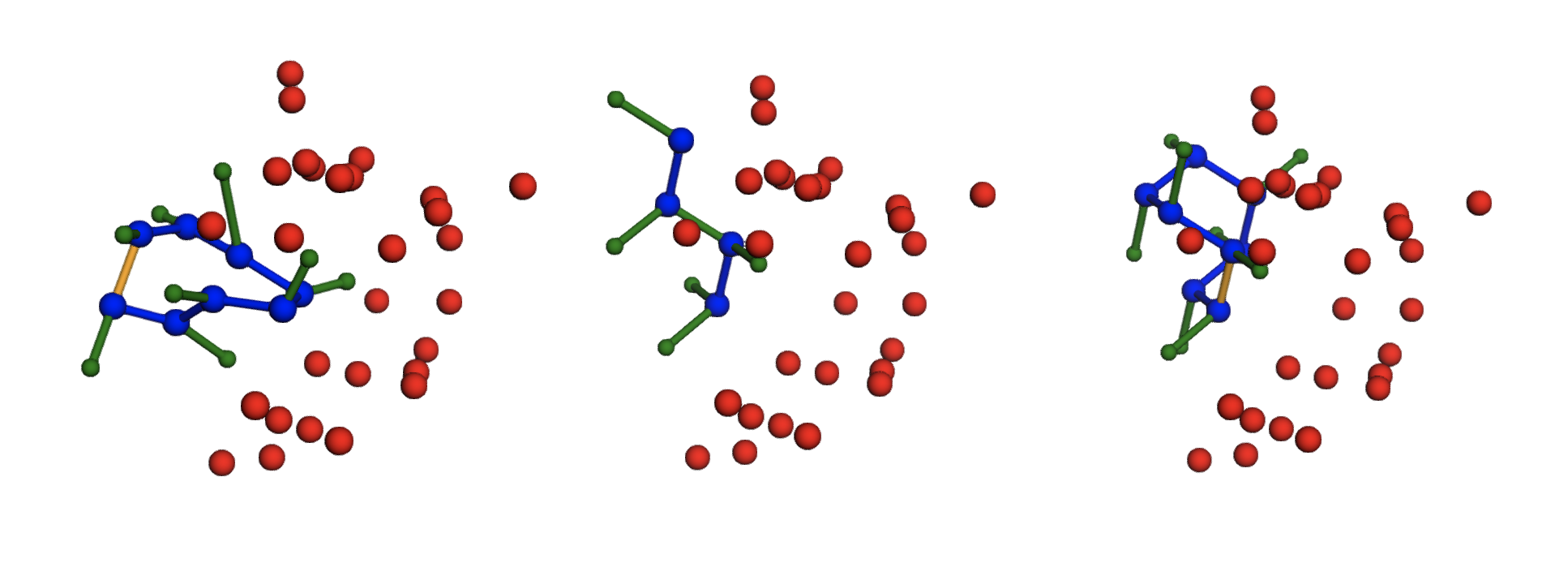}
\caption{3AV9 result visualization. True PDB peptide (left) against the QUBO model result (middle) and the CP-MJ 1-NN result (right). The peptide is represented with blue and green dots (main and side-chain, respectively, connected by lines), and the red dots are the external protein residues.}
\label{fig:3av9-result-comparison-plot} 
\end{figure}

\begin{figure}[ht!]
\centering
\includegraphics[width=0.9\textwidth]{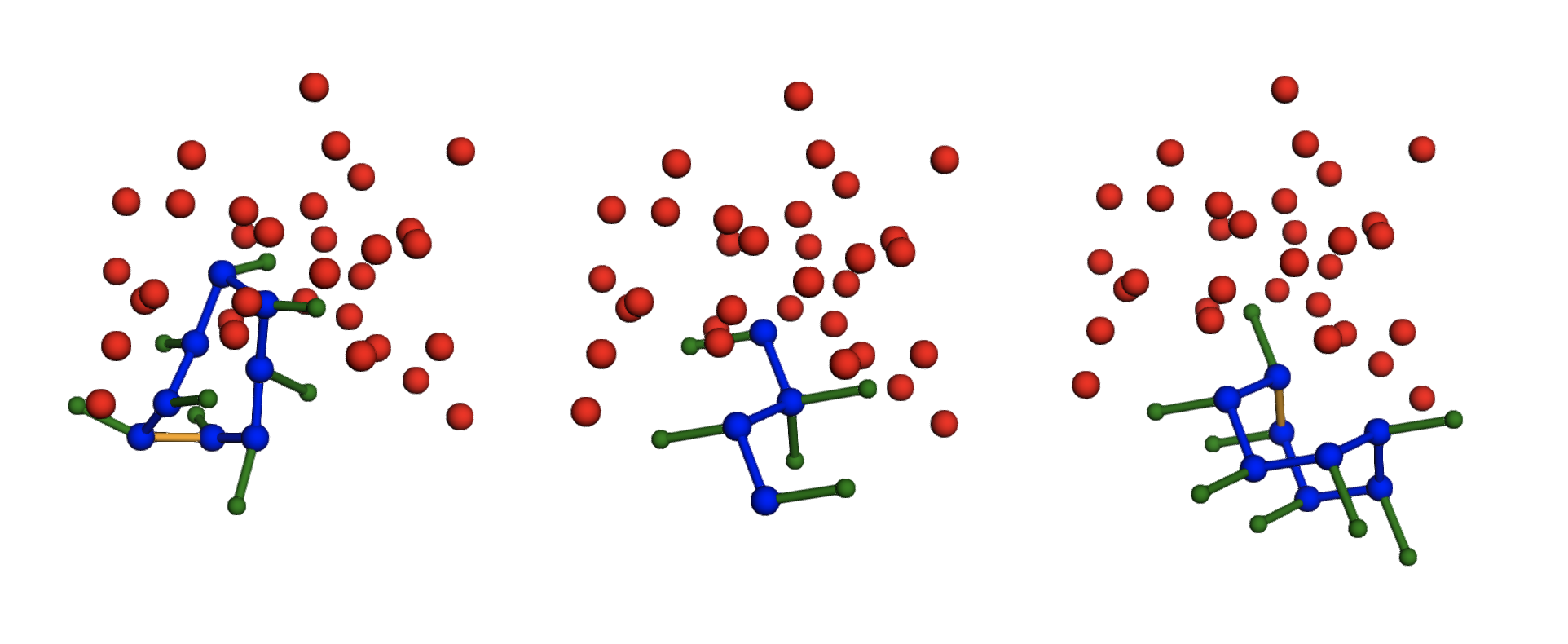}
\caption{3AVI result visualization. True PDB peptide (left) against the QUBO model result (middle) and the CP-MJ 1-NN result (right). The peptide is represented with blue and green dots (main and side-chain, respectively, connected by lines), and the red dots are the external protein residues.}
\label{fig:3avi-result-comparison-plot} 
\end{figure}

\begin{figure}[ht!]
\centering
\includegraphics[width=0.9\textwidth]{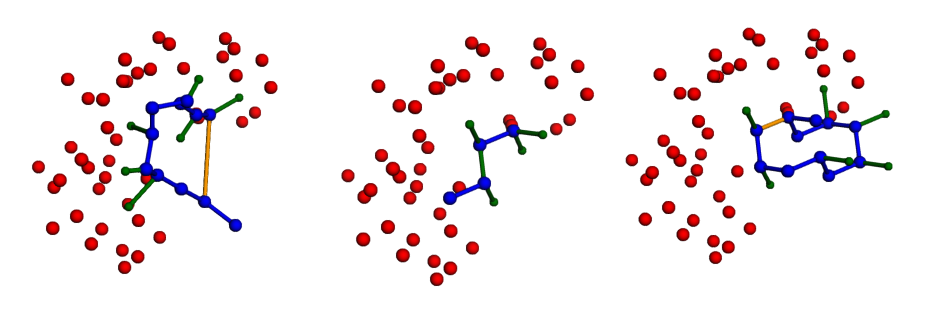}
\caption{2F58 result visualization. True PDB peptide (left) against the QUBO model result (middle) and the CP-MJ 1-NN result (right). The peptide is represented with blue and green dots (main and side-chain, respectively, connected by lines), and the red dots are the external protein residues.}
\label{fig:2f58-result-comparison-plot} 
\end{figure}

\section{Additional discussion points}\label{appendix:discussion}
\subsection{Construction of the Hamiltonian}

Initially, we pursued different problem encodings in parallel: the spatial encoding (see Appendix \ref{appendix:spatial}), the turn ancilla encoding (see Appendix \ref{appendix:turn_ancilla}), and the variable efficient turn encoding (see Section \ref{sec:problem}).
We implemented feature-complete versions of each, including all problem constraints and the external protein.
Because the spatial encoding is not anchored to a specific starting point (e.g., the lattice origin), we expected it to be more robust and likely the more performant encoding scheme.
However, the Hamiltonian variables and terms scaling of the spatial encoding are prohibitive at the problem sizes we are interested in solving. 
In fact, the number of terms in the spatial encoding QUBO Hamiltonian is so large that attempts at manipulating the Hamiltonian expression  overloaded our hardware and caused the Python process to terminate.
Because of this, we shifted our attention to one of the turn encoding approaches.

The turn ancilla approach also suffers from a scaling problem, namely, every distance calculation between pairs of residues must be upper-bounded by a fixed value, duplicated for every distance comparison] and this bound grows as the problem size grows. 
This means that the variable scaling of the turn ancilla approach 
scales worse than $\mathcal{O}(N^2)$ and also scales with a factor of $\log_2(N)$ [or $\log_2(\max(d(M,N)))$ considering the protein].
This is not true for the resource-efficient approach, which does not seek to bound all distance calculations, but rather includes ancilla bits to turn off the effect of those distances when necessary.
Because of this, we expect the resource-efficient approach to actually be the most efficient and therefore most useful approach in solving the docking problem, as demonstrated in the scaling experiments. 
Details on the Hamiltonian variable and terms scaling experiments can be found in Appendix \ref{appendix:scaling}.

\subsection{Lattice Considerations}

The choice of the tetrahedral lattice over higher-resolution alternatives such as cubic or face-centered cubic helped keep the problem tractable, and helped reduced problem complexity for the turn encodings. 
Cubic lattices have 6 turns from each vertex, which would require 3 bits and generate two invalid turn directions that need to be accounted for, and the face-centered cubic lattice has 4 turns from any given face vertex and 6 turns from a given corner vertex, so the representation would need to vary, or additional invalid turn directions would need to be accounted for.
In contrast, any vertex on the tetrahedral lattice has exactly 4 turns allowed from it, though they alternate in directions.
This means we could fix the representation at 2 bits per turn, and simply introduce a sign flip according to the specific sub-lattice.
This helps reduce the conceptual complexity, and it kept the Hamiltonian scaling to a minimum, whereas the cubic or face-centered cubic lattices would have exhibited even worse scaling.

Once the lattice structure is fixed, the depth (e.g., size) of the lattice needs to be established.
There is no rule for this, other than it needs to be big enough to fit the peptide and a sufficient amount of the protein to properly guide the peptide conformation.
It might seem that the lattice should be set big enough to fit the full protein on it---which is to say the full set of protein projections---but this is not necessarily the case, and may result in a lattice that is too large.
If the lattice is grown too large, a couple risks are introduced.
First, because we use the protein active site only, and not the full protein, regions around the sides of the protein (active site) start to appear, and any spatial encoding model may choose to place the peptide in these regions (assuming enough space exists).
Although we did not observe this, it is possible that the peptide could be in a region that would actually be inside of the full protein complex.
Second, and perhaps more importantly, allowing for full protein projection effectively doubles the number of protein influenced vertices that need to be included in the Hamiltonian.
This means that from a starting point near the active site centroid (i.e., the lattice origin), any peptide conformation would need to walk through a double layer of protein blocking sites before it reaches any interaction sites that are on the far side of the protein (with respect to the centroid).
As long as the protein clash penalty ($\mu_2$) is set high enough, any turn encoding model would never choose such a solution.
Similarly, for a spatial encoding if the lattice is set large enough to include those interaction projections, but no larger, then there would be no way for the spatial model to place the peptide in a way to reach those projections without also landing on the blocking sites.
As such, it does not help to include those interaction sites on the far side of the protein, as they introduce an additional scaling burden for all encodings.

Finally, we note that the use of a fixed lattice structure in our problem construction introduces some minimum error in the RMSD metric that cannot be overcome.
The lattice assumes a fixed set of turn directions and a fixed length between vertices, neither of which occurs in the real world.
This leads to model results that appear clean and highly structured, but that do not align with the more irregular conformations found in the PDB results.
Figures \ref{fig:3avn-result-comparison-plot} and \ref{fig:5lso-result-comparison-plot} show the differences clearly.
Given our problem definition, we find it more useful to focus on finding feasible solutions, and optimizing for high quality results per MJ interaction potentials within the feasible set.

\subsection{Looking forward}

While we stand by our work here as an important synthesis of prior art, and a valuable step forward in formulating the problem via the inclusion of the external protein, we have noted several limitations incurred by our simplifying assumptions. There are several alternatives for refinement.
First, we suggest revisiting the tetrahedral lattice geometry.
The tetrahedral geometry stemmed from the realistic inter-bond and dihedral angles commonly observed in atomistic space but it is not obvious if this property persists in two-particle CG amino acid representation. 
Continuous dihedral angle and variable bond lengths would be ideal but such implementation would be even harder to investigate using QUBO formulations. 
Second, we might consider revisiting the MJ interaction energies used to guide the conformation. 
MJ potentials are designed to capture effects between entire amino acids, and no CG equivalent potential exists in the literature to better represent the interaction potential for two-particle amino acids.
Third, we could extend the constraints to also include specifying chirality, which has real-world implications on the final conformation.
Fourth, the QUBO model currently only allows for 1-NN interactions.
However, the preferred interaction radius is approximately $6.5\AA$, which translates to roughly 2-NN interactions on the tetrahedral lattice.
Thus, an obvious extension would be to include 2-NN interactions between peptide residues.
Finally, we note that the current lattice structure is grown at a fixed location and angle relative to the protein residues.
There is no guarantee that this is the optimal lattice orientation, and in fact it is likely not to be the optimal orientation. 
It would be worthwhile to investigate methodologies to optimize lattice orientation, either as part of the solution pipeline or model solving, such that RMSD is further minimized.

\end{appendices}

\end{document}